\pdfoutput=1
\newif\ifjournal
\journalfalse 
\ifjournal
 
\documentclass[envcountsect,envcountsame,twocolumn,final]{svjour3}
\journalname{Natural Computing}
\else
\documentclass{article}
\fi

\usepackage{amsmath}
\usepackage{amssymb}
\usepackage{etoolbox}
\usepackage[utf8]{inputenc} 
\usepackage[T1]{fontenc} 
\usepackage{lmodern} 
\usepackage{tikz}
\usepackage{authblk}
\usepackage{physics}
\usetikzlibrary{positioning}
\usetikzlibrary{arrows.meta}
\usepackage{extdash}
\usepackage{hyperref}

\newif\ifbiblatex
\biblatexfalse
\ifbiblatex
\usepackage[backend=biber,maxbibnames=9,giveninits=true,isbn=false,mincrossrefs=999]{biblatex}
\AtEveryBibitem{\clearfield{number}}

\DeclareFieldFormat{titlecase}{\MakeTitleCase{#1}}
\newrobustcmd{\MakeTitleCase}[1]{
  \ifthenelse{\ifcurrentfield{booktitle}\OR\ifcurrentfield{booksubtitle}
    \OR\ifcurrentfield{maintitle}\OR\ifcurrentfield{mainsubtitle}
    \OR\ifcurrentfield{journaltitle}\OR\ifcurrentfield{journalsubtitle}
    \OR\ifcurrentfield{issuetitle}\OR\ifcurrentfield{issuesubtitle}
    \OR\ifentrytype{book}\OR\ifentrytype{mvbook}\OR\ifentrytype{bookinbook}
    \OR\ifentrytype{booklet}\OR\ifentrytype{suppbook}
    \OR\ifentrytype{collection}\OR\ifentrytype{mvcollection}
    \OR\ifentrytype{suppcollection}\OR\ifentrytype{manual}
    \OR\ifentrytype{periodical}\OR\ifentrytype{suppperiodical}
    \OR\ifentrytype{proceedings}\OR\ifentrytype{mvproceedings}
    \OR\ifentrytype{reference}\OR\ifentrytype{mvreference}
    \OR\ifentrytype{report}\OR\ifentrytype{thesis}}
    {#1}
    {\MakeSentenceCase{#1}}}

\DeclareSourcemap{
  \maps[datatype=bibtex]{
    \map{
      \step[fieldsource=doi,final]
      \step[fieldset=url,null]
    }
    \map[overwrite]{ 
      \step[fieldsource=shortjournal,fieldtarget=journaltitle]
    }
  }
}
\else
\fi

\ifjournal\else
\usepackage{amsthm}
\newtheorem{theorem}{Theorem}[section]

\newtheorem{remark}[theorem]{Remark}
\newtheorem{definition}[theorem]{Definition}
\newtheorem{example}[theorem]{Example}
\fi

\newcommand{\calN}{\mathcal{N}}
\newcommand{\calD}{\mathcal{D}}
\newcommand{\calC}{\mathcal{C}}
\newcommand{\calL}{\mathcal{L}}
\newcommand{\calT}{\mathcal{T}}

\DeclareMathOperator{\pre}{pre}
\DeclareMathOperator{\post}{post}
\newcommand{\N}{\mathbb{N}}
\newcommand{\R}{\mathbb{R}}
\newcommand{\Z}{\mathbb{Z}}

\renewcommand*{\vec}[1]{\mathbf{#1}}
\newcommand{\vi}{{\vec{i}}}
\newcommand{\vc}{{\vec{c}}}
\newcommand{\vd}{{\vec{d}}}
\newcommand{\vv}{{\vec{v}}}
\newcommand{\vr}{{\vec{r}}}
\newcommand{\vp}{{\vec{p}}}
\newcommand{\vo}{{\vec{o}}}
\newcommand{\vx}{{\vec{x}}}
\newcommand{\vy}{{\vec{y}}}

\newcommand{\reachone}{\Rightarrow}
\newcommand{\reach}{\Rightarrow^*}

\renewcommand{\emptyset}{\varnothing}
\newcommand*\chem[1]{\ensuremath{\mathrm{#1}}}

\title{Computing with Chemical Reaction Networks:\\ A Tutorial}
\author{Robert Brijder}
\ifjournal
\institute{Hasselt University, Belgium. \email{robert.brijder@uhasselt.be}}
\else
\affil{Hasselt University, Belgium}
\date{}
\fi

\begin{document}

\maketitle

\begin{abstract}
Chemical reaction networks (CRNs) model the behavior of chemical reactions in well-mixed solutions and they can be designed to perform computations. In this tutorial we give an overview of various computational models for CRNs. Moreover, we discuss a method to implement arbitrary (abstract) CRNs in a test tube using DNA. Finally, we discuss relationships between CRNs and other models of computation.
\end{abstract}

\section{Introduction}
Chemical reaction networks are a fundamental model of chemical reactions in well-mixed solutions. A \emph{chemical reaction network} (CRN) is, roughly, a finite set of chemical reactions like $X + Y \to 2Y + Z$, where $X$, $Y$, and $Z$ are ``abstract'' molecular species, i.e., these species are not tied to any chemical implementation. CRN theory \cite{Feinberg/CRNLectures}, which studies the dynamic behavior of CRNs, is a mature research field that is traditionally focused on networks of chemical reactions occurring in nature. Recently, it has been been shown that carefully designed CRNs are able to compute \cite{DBLP:conf/ecal/LiekensF07,DBLP:journals/nc/SoloveichikCWB08} --- such computations can, e.g., take place in a test tube. There is now a rapidly growing body of literature devoted to computational models for (abstract) CRNs. Decoupling reactions from chemical implementations introduces a higher level of abstraction, where CRNs become a high-level \emph{programming language}. Using a (semi\Hyphdash*)automated method, arbitrary abstract CRNs can then be \emph{compiled} to chemical implementations \cite{SimCRN/Soloveichik,ChenDSPCSS/CRNprog/NatureNano,BadeltSJDTW17/LNCS/Nuskell}. 

In this tutorial we give a gentle overview of the popular computational models for CRNs, a possible chemical implementation for arbitrary CRNs, and an overview of related models of computation.

First we define the notion of a CRN and the way it operates on a discrete state space (Section~\ref{sec:CRNs}). A discrete state describes the counts of the species of a CRN\@. In Subsections~\ref{ssec:comp_CRNs_deciding} and \ref{ssec:stably_deciding} we turn to a model of computation for such discrete CRNs inspired by the notion of population protocols from the research field of distributed computing \cite{DBLP:journals/dc/AngluinADFP06}. Here a CRN computes by either accepting or rejecting an input state, much like a finite state automaton or a Turing machine accepts or rejects arbitrary input strings. Equivalently, such a CRN can be seen as recognizing a Boolean-valued function on its set of input states. This computational model is then extended in Subsection~\ref{ssec:CRC} to the computation of more general functions than Boolean-valued functions \cite{DBLP:journals/nc/ChenDS14}. Next we study CRNs that never need ``slow'' reactions to perform their computations (such slow reactions are called speed faults) \cite{SpeedFaults/DC/CCDS/2017}. 

In Subsection~\ref{ssec:stoch_CRNs}, we augment discrete CRNs with stochastics to obtain stochastic CRNs. In this way, a stochastic CRN behaves as a continuous-time Markov chain on the discrete state space of the CRN\@. The probability for a reaction to take place here depends on (1) the molecular counts of the reactants, (2) the volume of the solution (e.g., the test tube), and (3) the rate constant (a value that depends on the reaction). With the notion of a stochastic CRN in place, we show in Subsection~\ref{ssec:SCRN_Turing_universal} that stochastic CRNs can simulate Turing machines with an arbitrary small positive probability of error \cite{DBLP:journals/nc/SoloveichikCWB08}. In Subsection~\ref{ssec:leader_election} we turn to the natural problem of leader election \cite{AngluinAE08a/DISTJournal/FastWithLeader,BellevilleDS17/ICALP/LeaderlessPopProt,DotyS/DISTJournal/LeaderElectionLinTime} and in Subsection~\ref{ssec:comp_prob_dist} we show that probability distributions can be computed using CRNs \cite{CardelliKL16/DNA22/ProgProbDistr}.

Instead of considering discrete state spaces, where a discrete state describes the \emph{counts} (which are nonnegative integers) of the species of a CRN, one can also consider a continuous state space, where a continuous state describes the \emph{concentrations} (which are nonnegative real numbers) of the species. In fact, CRN theory has traditionally focused largely on CRNs with continuous state spaces. However, such continuous CRNs only form a faithful approximation of reality in environments where the molecule counts are high (and stay high) \cite{Kurtz/ContLimitDiscrete}. The standard mass-action continuous CRN model is given in Subsection~\ref{ssec:ContinuousCRNs}, including a discussion of the computational mode of operation for this CRN model \cite{DBLP:conf/cmsb/FagesGBP17}, and the continuous CRN model from \cite{DBLP:conf/innovations/ChenDS14} is studied from a computational point of view in Subsection~\ref{ssec:rateIndepCRNcomp}.

In Section~\ref{sec:implement_DNAsd} we recall from \cite{SimCRN/Soloveichik} that arbitrary CRNs can be implemented in the wetlab using DNA as a substrate, in this way, justifying the high level of abstraction that was taken in the previous sections. We recall in Section~\ref{sec:related_fields} that CRNs are closely related to Petri nets \cite{PetriNet/review/Pet1977,DBLP:conf/ac/1996petri1}, vector addition systems \cite{VASsKarpMiller}, and population protocols \cite{DBLP:journals/dc/AngluinADFP06}. Petri nets and vector addition systems are the most studied models of concurrency and population protocols form a popular model for distributed computing. We end with a discussion.

\section{Chemical reaction networks}\label{sec:CRNs}
In this section we recall the notion of a chemical reaction network (CRN), which is roughly a set of reactions. We consider a level of abstraction that is higher than that of concrete chemical reactions. So, rather than considering ``concrete'' chemical reactions such as $\chem{NaHCO_3} + \chem{HCl} \to \chem{H_2O} + \chem{NaCl} + \chem{CO_2}$, we abstract from the level of molecular species, like $\chem{NaHCO_3}$ and $\chem{HCl}$, and instead consider abstract species, usually denoted by capital letters like $A$ and $B$, and reactions like $2A + B \to 3A$. Thus we do not worry about the chemical implementations of species $A$ and $B$. This higher level of abstraction is justified in Section~\ref{sec:implement_DNAsd}, where it is recalled that \emph{any} (abstract) CRN is implementable in the wetlab using DNA as a substrate.

Let $\N$ be the set of nonnegative integers. Let $\Lambda$ be a finite set. The set of vectors over $\N$ indexed by $\Lambda$ (i.e., the set of functions $\varphi: \Lambda \rightarrow \N$) is denoted by $\N^\Lambda$. A vector $\vv \in \N^\Lambda$ can be viewed as a multiset with $\Lambda$ as the underlying set. For $\vx \in \N^\Lambda$, we denote the cardinality of the multiset $\vx$ by $\| \vx \| = \sum_{i \in \Lambda} \vx(i)$. We denote the restriction of $\vx$ to $\Sigma \subseteq \Lambda$ by $\vx|_{\Sigma}$. For $\vx,\vy \in \N^\Lambda$ we write $\vx \leq \vy$ if and only if $\vx(i) \leq \vy(i)$ for all $i \in \Lambda$.

A \emph{reaction} $\alpha$ over $\Lambda$ is a tuple $(\vr,\vp)$ with $\vr,\vp \in \N^\Lambda$; $\vr$ and $\vp$ are called the \emph{reactants} and \emph{products} of $\alpha$, respectively. A reaction is commonly written additively, where, e.g., $A + 2B \to B + C$ denotes the reaction $(\vr,\vp)$ over $\Lambda = \{A,B,C\}$, where $\vr(A) = 1$, $\vr(B) = 2$, $\vr(C) = 0$, $\vp(A) = 0$, $\vp(B) = 1$, and $\vp(C) = 1$. Reaction $\alpha$ is called \emph{unimolecular} if $\| \vr \| = 1$, and \emph{bimolecular} if $\| \vr \| = 2$. Since it is very rare in nature for three or more molecules to simultaneously collide, almost all elementary reactions (that is, reactions that cannot be decomposed into multiple reactions) in nature are unimolecular and bimolecular reactions. Reaction $\alpha$ is called \emph{mute} if $\vr = \vp$.

We now define the central notion of a chemical reaction network.
\begin{definition}\label{def:CRN}
A \emph{chemical reaction network (CRN)} is an ordered pair $\calN = (\Lambda, R)$ with $\Lambda$ a finite set and $R$ a finite set of reactions over $\Lambda$.
\end{definition}

The elements of $\Lambda$ are called the \emph{species} of $\calN$. Also, the sets of species and reactions of a CRN are denoted by $\Lambda(\calN)$ and $R(\calN)$, or, if the CRN under consideration is clear, simply by $\Lambda$ and $R$, respectively.

\begin{example}\label{ex:CRN}
Let $\calN = (\Lambda, R)$ with $\Lambda = \{ A, B, C \}$ and $R = \{ 3A \to 2B, B + C \to A, C \to B, B \to C \}$. Then $\calN$ is a CRN having three species and four reactions. One reaction of $\calN$ is bimolecular ($B + C \to A$) and two are unimolecular ($C \to B$ and $B \to C$).
\end{example}

The elements of $\N^\Lambda$ are called the (discrete) \emph{states} of $\calN$ (also called configurations in the literature), and they describe the counts of each of the molecular species of $\calN$ in some well-mixed solution (such as a well-mixed test tube). Viewing $\vc$ as a multiset, each element of $\vc$ is called a \emph{molecule}. So, $\vc$ has $\| \vc \|$ molecules. A molecule of species $S$ is sometimes called a $S$-molecule for short. Just as reactions, we often write states additively (assuming the underlying species set $\Lambda$ is clear from the context). If $S \in \Lambda$ is some species and $\vc$ a state then the number $\vc(S)$ of $S$-molecules in $\vc$ is sometimes denoted by $\#_\vc S$ or simply $\# S$ if $\vc$ is clear from the context.

As a consequence of the well-mixedness assumption, if all reactants of some reaction $\alpha = (\vr,\vp)$ are available in sufficient quantity (i.e., $\vr \leq \vc$), then $\alpha$ can take place. This is formalized as follows.

For a state $\vc \in \N^\Lambda$ and a reaction $\alpha$ over $\Lambda$, we say that $\alpha = (\vr,\vp)$ is \emph{applicable} to $\vc$ if $\vr \leq \vc$. If $\alpha$ is applicable to $\vc$, then the \emph{result} of applying $\alpha$ to $\vc$, denoted by $\alpha(\vc)$, is $\vc' = \vc-\vr+\vp$. In this case we also write $\vc \reachone_\alpha \vc'$. Note that $\vc'$ is a state, i.e., $\vc' \in \N^\Lambda$. We also write $\vc \reachone_\calN \vc'$ to denote that $\vc \reachone_\alpha \vc'$ for some reaction $\alpha$ of $\calN$. The transitive and reflexive closure of $\reachone_{\calN}$ is denoted by $\reach_{\calN}$. If $\vc \reach_\calN \vc'$, then we say $\vc'$ is \emph{reachable} from $\vc$ in $\calN$.

\begin{example}
Consider again the CRN $\calN$ of Example~\ref{ex:CRN}. Consider the state $\vc = A + 2C$. Then only the reaction $C \to B$ of $\calN$ is applicable to $\vc$. We have $\vc \reachone_\calN \vc'$ where $\vc' = A + B + C$, in other words $\vc'$ is the result of applying $C \to B$ to $\vc$. Three reactions of $\calN$ are applicable to $\vc'$. For example, we have $\vc' \reachone_\calN 2A$. In state $2A$ no reactions of $\calN$ are applicable. We observe that, e.g., $2A$ is reachable from $\vc$.
\end{example}

\begin{remark}
We remark that a reaction $\alpha$ is usually defined as a triple, consisting also of a positive real number $k_\alpha$ called the rate constant of $\alpha$ which determines the likelihood of the reaction to take place in the current state (assuming it is applicable). Since this section and the next section only deals with reachability (i.e., whether it is \emph{possible} to reach one state from another), we postpone considering rate constants until Section~\ref{sec:comp_stoch_CRNs}.
\end{remark}

For $\vc \in \N^\Lambda$, we define $\pre_{\calN}(\vc) = \{ \vc' \in \N^\Lambda \mid \vc' \reach_\calN \vc \}$ and $\post_{\calN}(\vc) = \{ \vc' \in \N^\Lambda \mid \vc \reach_\calN \vc' \}$. So, $\post_{\calN}(\vc)$ contains all states that can be reached from $\vc$ (including $\vc$ itself), and $\pre_{\calN}(\vc)$ contains all states that can reach $\vc$ (including $\vc$ itself). A state $\vc \in \N^\Lambda$ is called \emph{terminal} in $\calN$ if $\post_{\calN}(\vc) = \{\vc\}$. In other words, a state is terminal if no non-mute reaction of $\calN$ is applicable to $\vc$.

If $\calN$ is clear from the context, then we often omit the subscripts of $\reachone_{\calN}$, $\reach_{\calN}$, $\pre_{\calN}$ and $\post_{\calN}$.

We extend $\pre(\vc)$ and $\post(\vc)$ to sets $X \subseteq \N^\Lambda$ of states in the natural way: $\pre(X) := \bigcup_{\vc \in X} \pre(\vc)$, and $\post(X) := \bigcup_{\vc \in X} \post(\vc)$.

We remark here that the notion of a CRN is similar to some notions from other research domains, see Section~\ref{sec:related_fields} for details. Therefore, the results presented here can (often) be straightforwardly carried over to these domains.

\section{Computing with discrete chemical reaction networks}\label{sec:comp_CRNs}
For a significant part of the rest of the paper we recall several models of computing with CRNs from the literature and discuss some of their key results. The computational CRN models this only concern reachability of states, and so their results are independent of stochastics (i.e., how likely a certain state is reached). CRNs augmented with stochastics are discussed in Section~\ref{sec:comp_stoch_CRNs}.

For didactical reasons we do not discuss the computational CRN models in chronological order, but instead we first consider the elementary computational model introduced in \cite{DBLP:journals/nc/ChenDS14}, which is in turn inspired by (and very similar to) the computational model of Population Protocols \cite{DBLP:journals/dc/AngluinADFP06} (see Subsection~\ref{ssec:PopProt} for a comparison between Population Protocols and CRNs).

\subsection{Haltingly deciding chemical reaction deciders}\label{ssec:comp_CRNs_deciding}
Suppose we are given a state with an unknown number of molecules of species $X$ and $Y$ and we want to decide whether or not $\# X$ is equal to $\# Y$ modulo $3$.
Is there a CRN that can perform this computation? More specifically, is there a CRN which eventually (by keeping applying reactions) halts on every possible input such that merely the presence of certain molecules in the halting state indicates whether the answer to the decision problem is yes or no? The next example (taken from \cite{BrijderDS/CompModes/NatComp}) shows that the above given modulo problem $\# X \stackrel{?}{\equiv} \#Y \mod 3$ can be decided by a CRN.

\begin{example}\label{ex:mod3_compute}
Consider the CRN $\calN = (\Lambda, R)$ with $\Lambda = \{ X, Y, V \}$ and $R$ consisting of the following reactions
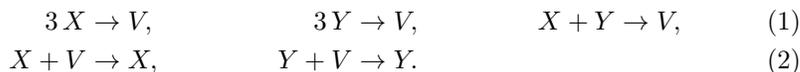
\begin{align}
3\,X &\to V, & 3\,Y &\to V, & X + Y &\to V, \label{eqn:eatXY}\\
X + V &\to X, & Y + V &\to Y. \label{eqn:eatV}
\end{align}
First notice that all reactions preserve whether or not $\# X \equiv \#Y \mod 3$. The reactions of (\ref{eqn:eatXY}) all reduce the number of $X$ or $Y$ molecules, while the remaining reactions (of (\ref{eqn:eatV})) do not influence the $X$ and $Y$ molecules. So, eventually (that is, when reactions continue to take place), we reach a state where none of the reactions of (\ref{eqn:eatXY}) can take place anymore. The last reaction of (\ref{eqn:eatXY}) that took place introduced a $V$-molecule. Now, if $\# X \equiv \#Y \mod 3$, then no $X$ and $Y$ molecules are present anymore at this point and so the CRN has halted with only some $V$-molecules. If $\# X \not\equiv \#Y \mod 3$, then some $X$- or $Y$-molecules remain and these eat all the $V$-molecules that are present by the reactions of (\ref{eqn:eatV}). So, in this case the CRN eventually halts with only some $X$- or $Y$-molecules.

Consequently, eventually the CRN halts and the presence of $V$-molecules in the terminal state indicate that $\# X \equiv \#Y \mod 3$ holds, while the presence of $X$- or $Y$-molecules in the terminal state indicate that $\# X \equiv \#Y \mod 3$ does not hold. We then say that $V$ is a \emph{yes voter} (or \emph{1-voter}) and $X$ and $Y$ are \emph{no voters} (or \emph{0-voters}). Note however that this computation does not work in the corner case where the initial state has no molecules, since in this case no $V$-molecule is ever produced.
\end{example}

Inspired by Example~\ref{ex:mod3_compute} we now formalize the above illustrated model of computation for CRNs. First we define the notion of chemical reaction decider which is roughly a CRN augmented with three distinguished sets of species: one to define the input states, one to define the no voters, and one to define the yes voters.

\begin{definition}
A \emph{chemical reaction decider (CRD)} is a 4-tuple $\calD = (\calN,\Sigma,\Lambda_0,\allowbreak \Lambda_1)$, where $\calN$ is a CRN, $\Sigma, \Lambda_0,\allowbreak \Lambda_1 \subseteq \Lambda(\calN)$, and $\Lambda_0 \cap \Lambda_1 = \emptyset$.
\end{definition}
The elements of $\Sigma$, $\Lambda_0$, and $\Lambda_1$ are called the \emph{input species}, \emph{0-voters}, and \emph{1-voters}, respectively. The elements of $\N^\Sigma \setminus \{\vec{0}_\Sigma\}$ are called the \emph{input states}, where by $\vec{0}_\Sigma$ we denote the zero vector with index set $\Sigma$. If the index set is clear from the context we just write $\vec{0}$ instead of $\vec{0}_\Sigma$. For $b \in \{0,1\}$, let $\calL_b = \{ \vc \in \N^\Lambda \mid c|_{\Lambda_b} \neq \vec{0} \}$ be the set of states that have at least one $b$-voter. We say that $\vc$ has \emph{output} $b \in \{0,1\}$ if $\vc \in \calL_b \setminus \calL_{1-b}$. In other words, $\vc$ has output $b$ when it contains $b$-voter molecules, but no $(1-b)$-voter molecules.

Let $\calT$ be the set of terminal states of $\calN$ and let, for $b \in \{0,1\}$, $\calT_b = \calT \cap (\calL_b \setminus \calL_{1-b})$ be the set of terminal states of $\calN$ with output $b$. We say that $\vc \in \N^\Lambda$ is \emph{output-$b$ halting} if $\post(\vc) \subseteq \pre(\calT_b)$. In other words, $\vc$ is output-$b$ halting if every state reachable from $\vc$ (including $\vc$ itself) can reach an output-$b$ terminal state. Note that a state cannot be both output-$0$ halting and output-$1$ halting.

\begin{remark}\label{rem:eventually}
It is worthwhile to note that the definition of output-$b$ halting is often sloppily interpreted as saying that starting from an output-$b$ halting state we \emph{eventually} reach an output-$b$ terminal state. This is incorrect in general since we may, e.g., have a output-$b$ halting (but not terminal) state $\vc$ such that $\vc \reachone^+ \vc$ (where $\reachone^+$ denotes the transitive closure of $\reachone$), and so the current state may indefinitely be in a loop without reaching an output-$b$ terminal state (we correctly used ``eventually'' in Example~\ref{ex:mod3_compute} since loops cannot appear there). To ensure eventually reaching an output-$b$ terminal state, it is possible to additionally assume some notion of \emph{fairness} \cite{DBLP:journals/nc/ChenDS14}. 
One such notion of fairness is implicit for stochastic chemical reaction networks (see Section~\ref{sec:comp_stoch_CRNs}) that have only a finite number of states reachable from any given state \cite{DotySoloveichik/ZeroError}.
\end{remark}

For $\vc \in \N^\Sigma$, let $\imath(\vc) \in \N^\Lambda$ be the vector obtained from $\vc$ by padding zeros for the entries indexed by $\Lambda \setminus \Sigma$, i.e., $\imath(\vc)|_{\Sigma} = \vc$ and $\imath(\vc)|_{\Lambda \setminus \Sigma} = \vec{0}$.

We now define a key notion.

\begin{definition}
We say that a CRD $\calD$ is \emph{haltingly deciding} if, for each input state $\vc$ of $\calD$, $\imath(\vc)$ is output-$b$ halting for some $b \in \{0,1\}$.
\end{definition}

Thus $\calD$ is haltingly deciding if for each input state $\vc$, there is a $b \in \{0,1\}$ such that during the computation a terminal state with output $b$ is always reachable. So, starting from $\vc$ you can never go ``wrong'' since a terminal state with output $b$ always remains reachable.

If $\calD$ is haltingly deciding, then we say that $\calD$ \emph{(haltingly) recognizes} the set $\{ \vc \in \N^\Sigma \setminus \{\vec{0}\} \mid \imath(\vc)$ is output-$1$ halting$\}$. If a haltingly deciding $\calD$ recognizes $X \subseteq \N^\Sigma \setminus \{\vec{0}\}$, then we also say that $\calD$ \emph{decides} the predicate, i.e., Boolean-valued function, $\varphi: \N^\Sigma \setminus \{\vec{0}\} \to \{0,1\}$ where $\varphi(x)$ holds (i.e., $\varphi(x) = 1$) if and only if $x \in X$.

\begin{example}
Consider again the CRN $\calN$ from Example~\ref{ex:mod3_compute}. With the notions and terminology in place we can now formalize the behavior of $\calN$ as a CRD $\calD = (\calN,\Sigma,\Lambda_0,\Lambda_1)$, where $\Sigma = \Lambda_0 = \{X,Y\}$ and $\Lambda_1 = \{V\}$. From the observations we made in Example~\ref{ex:mod3_compute} we conclude that $\calD$ haltingly recognizes the set $\{ \vc \in \N^\Sigma \setminus \{\vec{0}\} \mid \vc(X) \equiv \vc(Y) \mod 3 \}$.
\end{example}

\begin{remark}
Note that, when a CRD halts on a given input, the CRD does not give a ``signal'' that it has halted. In other words, an observer of a computation of the CRD does not know whether or not the output of the computation is final unless it has determined somehow that the computation has terminated (i.e., that no non-mute reaction can take place in the current state). One could imagine an alternative mode of operation for CRDs in which the presence of at least one molecule of some distinguished species signals that the output is final.
\end{remark}

It is natural to ask which sets can be haltingly recognized by CRDs. We say that $X \subseteq \N^\Sigma$ is \emph{linear} if there is a finite set $S \subseteq \N^\Sigma$ and a $\vd \in \N^\Sigma$ such that $X = \{\vd + \sum_{\vv \in S} n_\vv \vv \mid n_\vv \in \N \text{ for all } \vv \in S \}$. We say that $X \subseteq \N^\Sigma$ is \emph{semilinear} if $X$ is the union of a finite number of linear sets. We remark that semilinear sets are precisely the sets definable in Presburger arithmetic, which is the first-order theory of natural numbers with addition \cite{ginsburg1966}.

\newcommand{\xsizefig}{10}
\newcommand{\ysizefig}{6}

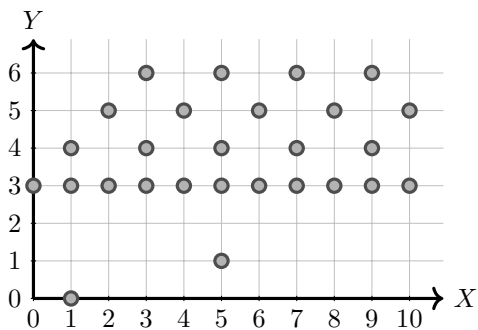
\begin{figure}
\begin{center}
\begin{tikzpicture}
[cfg/.style={circle,very thick,minimum size=5,inner sep=0,draw=black!70,fill=black!30},scale=0.5]
\draw[step=1,lightgray,very thin] (0,0) grid (\xsizefig.9,\ysizefig.9);
\draw [very thick,->] (0,0) -- (\xsizefig.9,0) node[right] {$X$};
\draw [very thick,->] (0,0) -- (0,\ysizefig.9) node[above] {$Y$};
\foreach \x/\xtext in {0,...,\xsizefig}
  \draw[shift={(\x,0)}] (0,2pt) -- (0,-2pt) node[below] {$\xtext$};

\foreach \y/\ytext in {0,...,\ysizefig}
  \draw[shift={(0,\y)}] (2pt,0) -- (-2pt,0) node[left] {$\ytext$};

\foreach \x in {0,...,\xsizefig}
  \node[cfg] at (\x,3) {};

\foreach \z in {0,...,4}
  \node[cfg] at (2*\z+1,4) {};

\foreach \z in {0,...,4}
  \node[cfg] at (2*\z+2,5) {};

\foreach \z in {0,...,3}
  \node[cfg] at (2*\z+3,6) {};

\node[cfg] at (1,0) {};
\node[cfg] at (5,1) {};

\end{tikzpicture}
\end{center}
\caption{A semilinear set.}
\label{fig:semilinear}
\end{figure}

\begin{example}
Let $\Sigma = \{X, Y\}$. Consider the following four linear sets: $L_1$ is defined by $S_1 = \{X+Y,2X\}$ and $\vd_1 = 3Y$, $L_2$ is defined by $S_2 = \emptyset$ and $\vd_2 = X$, $L_3$ is defined by $S_3 = \emptyset$ and $\vd_3 = 5X+Y$, and $L_4$ is defined by $S_4 = \{X\}$ and $\vd_4 = 3Y$. The semilinear set $L_1 \cup L_2 \cup L_3 \cup L_4$ is depicted in Figure~\ref{fig:semilinear}.
\end{example}

The following result combines results from \cite{DBLP:journals/dc/AngluinADFP06} and \cite{DBLP:journals/dc/AngluinAER07}. The if direction has been shown in the proof of the main result of \cite{DBLP:journals/dc/AngluinADFP06} (see \cite{NACO/Brijder16/outputUnstableConf} for some details concerning the halting claim), and the only-if direction is a special case of the main result of \cite{DBLP:journals/dc/AngluinAER07}.

\begin{theorem}[\cite{DBLP:journals/dc/AngluinADFP06,DBLP:journals/dc/AngluinAER07}]\label{thm:char_semilinear}
Let $X \subseteq \N^\Sigma \setminus \{\vec{0}\}$. Then $X$ is recognized by a haltingly deciding CRD if and only if $X$ is semilinear.

Moreover, this result also holds if we restrict to haltingly deciding CRDs $\calD$ such that (1) $\calD$ has only bimolecular reactions and (2) every species of $\calD$ is either a $0$-voter or a $1$-voter.
\end{theorem}

It is well known (from vector addition system theory) that for a state $\vc$ and a CRN $\calN$, $\pre_\calN(\vc)$ and $\post_\calN(\vc)$ are not necessarily semilinear \cite{tcs/Hopcroft/nonsemilinear}. This makes Theorem~\ref{thm:char_semilinear} rather surprising.

It is well known (see, e.g., \cite{ginsburg1966}) that semilinear sets are exactly the sets that are obtained by finite unions, intersections, and complementations of sets which are either of the form $X_{\vec{a},b} = \{ \vx \in \N^\Sigma \mid \vec{a} \cdot \vx \leq b\}$ or of the form $X_{\vec{a},b,m} = \{ \vx \in \N^\Sigma \mid \vec{a} \cdot \vx \equiv b \mod m \}$, where $\vec{a} \in \Z^\Sigma$, $b \in \Z$, and $m \in \N \setminus \{0,1\}$ are constants and $\cdot$ denotes the dot product. The proof of the if direction in \cite{DBLP:journals/dc/AngluinADFP06} shows that (1) there are CRDs that compute $X_{\vec{a},b}$ and $X_{\vec{a},b,m}$ and (2) if CRDs $\calD_1$ and $\calD_2$ compute the sets $X_1$ and $X_2$, respectively, then there are CRDs that compute $X_1 \cup X_2$, $X_1 \cap X_2$ and $\N^\Sigma \setminus X_1$.

As an example we illustrate why $X_{\vec{a},b,m}$ is semilinear.
\begin{example}\label{ex:modulo_compute_elaborate}
First we show that the predicate $\# Z \stackrel{?}{\equiv} b \mod m$, where $b$ and $m$ are nonnegative integers with $b < m$, can be haltingly decided by a CRD\@. Consider the CRD $\calD = (\calN, \Sigma, \Lambda_0, \Lambda_1)$ with $\calN = (\Lambda, R)$, $\Lambda = \{ Z, T, F, F_0 \}$, $\Lambda_0 = \{F_0,F\}$, $\Lambda_1 = \{T\}$ and $R$ consisting of the following reactions
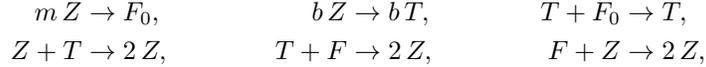
\begin{align*}
m\,Z &\to F_0, & b\,Z &\to b\,T, & T+F_0 &\to T,\\
Z + T &\to 2\,Z, & T + F &\to 2\,Z, & F + Z &\to 2\,Z,
\end{align*}
and $k\,Z \to k\,F$ for $k \in \{1,\ldots,m-1\} \setminus \{b\}$. Note that each of these reactions preserve the value $\#Z + \#F + \#T \mod m$. Because of the reaction $m\,Z\to F_0$ it is always possible to reach a state where $\#Z < m$. In fact, by additionally using the three reactions with $2\,Z$ as the products, it is always possible to reach a state where $x = \#Z + \#F + \#T < m$. If $x \neq 0$, then one easily verifies that the only way to halt is when the three reactions with $2\,Z$ as the products take place until $\#F + \#T = 0$, and then followed by either (1) reaction $b\,Z \to b\,T$ taking place (when $x=b$), or (2) $k\,Z \to k\,F$ taking place (when $x=k$). In case (2) or when $x=0$, the CRD has halted with only $0$-voters left, and in case (1) the CRN halts with only $T$-molecules left once $T+F_0 \to T$ has taken place repeatedly until all $F_0$-molecules are gone.

Now, more elaborate examples like $2\# X_1 - \# X_2 \stackrel{?}{\equiv} b \mod m$ can be reduced to the problem $\# Z \stackrel{?}{\equiv} b \mod m$ by extending the above $\calN$ with the reactions $X_1 \to 2\,Z$ and $X_2 \to (m-1)\,Z$ (the latter because $-1 \equiv m-1 \mod m$).
\end{example}

\subsection{Stably deciding chemical reaction deciders and other modes of operation}\label{ssec:stably_deciding}

We now present a natural generalization of the notion of haltingly deciding CRDs. Instead of requiring for each input the existence of some $b \in \{0,1\}$ such that during the computation a terminal state with output $b$ is always reachable, we now merely require that it is always possible to reach a state $\vc$ such that any state reachable from $\vc$ (including itself) has output $b$. So, even though non-mute reactions may still take place at $\vc$, any state reachable from $\vc$ has the same output $b$. State $\vc$ is then called output-$b$ stable.

More precisely, let $b \in \{0,1\}$. We say that $\vc \in \N^\Lambda$ is \emph{output-$b$ stable} if every $\vc' \in \post(\vc)$ has output $b$. Let $\mathcal{S}_b$ be the set of output-$b$ stable states. Note that any output-$b$ halting state is output-b stable, i.e., $\calT_b \subseteq \mathcal{S}_b$. 

Similar as for $\calT_b$, we say that $\vc \in \N^\Lambda$ is \emph{output-$b$ stabilizing} (for $b \in \{0,1\}$) if $\post(c) \subseteq \pre(\mathcal{S}_b)$. Note that a state cannot be both output-$0$ stabilizing and output-$1$ stabilizing. 

\begin{definition}
We say that a CRD $\calD$ is \emph{stably deciding} if for each input state $\vc$ of $\calD$, $\imath(\vc)$ is output-$b$ stabilizing for some $b \in \{0,1\}$.
\end{definition}

If $\calD$ is stably deciding, then we say that $\calD$ \emph{(stably) recognizes} the set $\{ \vc \in \N^\Sigma \setminus \{\vec{0}\} \mid \imath(\vc)$ is output-$1$ stabilizing$\}$.

While it is computationally easy to determine if a state is terminal (one just has to verify whether a non-mute reaction can take place in the given state), it does not seem to be computationally easy to determine if a state is output-$b$ stable for some $b \in \{0,1\}$ (although it is known to be decidable \cite{NACO/Brijder16/outputUnstableConf}). However, restricting to the class of CRDs where $\|\vr\| = \|\vp\| = 2$ for all reactions $\alpha = (\vr,\vp)$, output stability can be shown efficiently --- especially, due to a preprocessing step, when a large set of states need to be checked for output stability \cite{NACO/Brijder16/outputUnstableConf}.

The main result of \cite{DBLP:journals/dc/AngluinAER07} shows that a very general class of CRDs, which in particular includes the stably deciding CRDs, can only compute semilinear sets. By Theorem~\ref{thm:char_semilinear}, we therefore observe that stably deciding CRDs and haltingly deciding CRDs compute the same family of sets, namely the family of semilinear sets that do not contain the zero vector.

\begin{theorem}[\cite{DBLP:journals/dc/AngluinADFP06,DBLP:journals/dc/AngluinAER07}]\label{cor:CRD_semilinearity}
Let $X \subseteq \N^\Sigma$. Then $X$ is recognized by a stably deciding CRD if and only if $X$ is recognized by a haltingly deciding CRD.
\end{theorem}

Note that the above definitions of the output of a state are based on consensus: states with both yes and no voters do not have a defined output. One can consider a democratic mode of operation based on majority voting. Also, one can consider a mode of operation without $0$-voters (here the existence or absence of $1$-voters determines the output). These and other modes of operation have been shown to also compute exactly all semilinear sets not containing the zero vector, see \cite{BrijderDS/CompModes/NatComp}.

\subsection{Computing functions}\label{ssec:CRC}
In the previous subsection we considered a way of computing predicates $\varphi: \N^\Sigma \setminus \{\vec{0}\} \to \{0,1\}$ using CRNs. Following \cite{DBLP:journals/nc/ChenDS14}, we now consider the computation of functions of the form $\varphi: \N^\Sigma \to \N^\Gamma$ using CRNs.

\begin{example}\label{ex:minCRC}
Consider the function $\min$ that computes the minimum $\min(x,y)$ of two nonnegative integers $x$ and $y$.
For $\Sigma = \{X,Y\}$ and $\Gamma = \{Z\}$, the function $\min: \N^\Sigma \to \N^\Gamma$ can be easily seen to be computed through the reaction $X + Y \to Z$. Indeed, starting from an initial state $\vi$ consisting of only $X$ and $Y$ molecules, the CRN eventually halts in a state $\vc$ where $\#_{\vc} Z = \min(\#_{\vi}X, \#_{\vi}Y)$.
\end{example}

Computing the $\max$ function turns out to be slightly more involved.
\begin{example}\label{ex:maxCRC}
Consider the function $\max$ that computes the maximum $\max(x,y)$ of two nonnegative integers $x$ and $y$. For $\Sigma = \{X,Y\}$ and $\Gamma = \{Z\}$, the function $\max: \N^\Sigma \to \N^\Gamma$ can be computed using the auxiliary species $X'$ and $Y'$ and reactions 
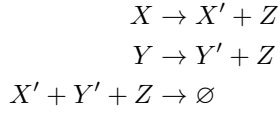
\begin{align*}
X &\to X' + Z \\
Y &\to Y' + Z \\
X' + Y' + Z &\to \emptyset
\end{align*}
Notice that if the third reaction is absent, then, starting from an initial state $\vi$ consisting of only $X$ and $Y$ molecules, the CRN eventually halts in a state $\vc$ where $\#_{\vc} Z = \#_{\vi}X + \#_{\vi}Y$. Observe that the third reaction consumes exactly $\min(\#_{\vi}X, \#_{\vi}Y)$ $Z$-molecules. So, the whole CRN eventually halts in a state $\vc$ where $\#_{\vc} Z = \#_{\vi}X + \#_{\vi}Y - \min(\#_{\vi}X, \#_{\vi}Y) = \max(\#_{\vi}X, \#_{\vi}Y)$. So, this CRN indeed computes the $\max$ function.
\end{example}

In analogy with the chemical reaction decider we define the 
chemical reaction computer \cite{DBLP:journals/nc/ChenDS14,DBLP:journals/nc/DotyH15}.
\begin{definition}\label{def:CRC}
A \emph{chemical reaction computer (CRC)} is 3-tuple $\calC = (\calN,\Sigma,\Gamma)$, where $\calN$ is a CRN and $\Sigma,\Gamma \subseteq \Lambda(\calN)$ are disjoint.
\end{definition}

The elements of $\Sigma$ and $\Gamma$ are called the \emph{input species} and \emph{output species}, respectively. Similar as for CRDs, the elements of $\N^\Sigma$ are called the \emph{input states}. We say that a state $\vc$ has \emph{output} $\vc|_{\Gamma}$.

We say that $\vc \in \N^\Lambda$ is \emph{output stable} if every $\vc' \in \post(\vc)$ has the same output as $\vc$. 

For $\vo \in \N^\Gamma$, let $\mathcal{S}_\vo$ be the set of output stable states with output $\vo$. We say that $\vc \in \N^\Lambda$ is \emph{output-$\vo$ stabilizing} if $\post(c) \subseteq \pre(\mathcal{S}_\vo)$. Note that a state is output-$\vo$ stabilizing for at most one $\vo \in \N^\Gamma$.

\begin{definition}
We say that a CRC $\calC$ is \emph{stably computing} if for each input state $\vc$ of $\calC$, $\imath(\vc)$ is output-$\vo$ stabilizing for some $\vo \in \N^\Gamma$. 
\end{definition}

If $\calC$ is stable deciding, then we say that $\calC$ \emph{(stably) computes} the function $\varphi: \N^\Sigma \to \N^\Gamma$ where $\varphi(\vc) = \vo$ if $\vc$ is output-$\vo$ stabilizing.

Finally, $\varphi: \N^\Sigma \to \N^\Gamma$ is called \emph{semilinear} if the set $\{ \vc \in \N^{\Sigma \cup \Gamma} \mid \varphi(\vc|_{\Sigma}) = \vc|_{\Gamma} \}$ is semilinear.

\begin{example}
Consider again $\min: \N^\Sigma \to \N^\Gamma$ from Example~\ref{ex:minCRC}. We can easily verify that $\min$ is semilinear. Indeed $\min(x,y)=z$ if and only if $(y = z \land x \geq z) \lor (x = z \land y \geq z)$. Since $y=z$ is equivalent to $y \leq z \land z \leq y$, we observe that $\{ (x,y,z) \mid \min(x,y)=z \}$ can be expressed by finite unions and intersections of sets of the form $X_{\vec{a},b} = \{ \vx \in \N^\Sigma \mid \vec{a} \cdot \vx \leq b\}$. Indeed, e.g., $y \leq z$ is expressed as $X_{(0,1,-1),0}$. Thus $\min$ is semilinear. In the same way we observe that $\max$ is semilinear.
\end{example}

The following is shown in \cite{DBLP:journals/nc/DotyH15} (by using results from \cite{DBLP:journals/dc/AngluinADFP06,DBLP:journals/dc/AngluinAER07,DBLP:journals/nc/ChenDS14}).
\begin{theorem}[\cite{DBLP:journals/nc/DotyH15}]\label{thm:CRC_semilinearity}
Stably computing CRCs compute exactly the semilinear functions $\varphi: \N^\Sigma \to \N^\Gamma$ with $\varphi(\vec{0})=\vec{0}$.
\end{theorem}

\subsection{Speed faults}\label{ssec:speed_faults}
A reaction $\alpha$ is considered ``slow'' in a state $\vc$ if at least two reactants of $\alpha$ appear in low quantity in $\vc$. We will see in Subsection~\ref{ssec:stoch_CRNs} that, assuming the standard stochastic CRN model, the expected time of such reactions to take place is indeed long.

Let $\alpha$ be a uni- or bimolecular reaction. Then $\alpha$ is called \emph{$k$-fast} for $\vc$, denoted by $\vc \reachone_{\alpha,\geq k} \vc'$, if $\#_{\vc}X \geq k$ for some reactant $X$ of $\alpha$. Similarly as before, we define $\reachone_{\calN,\geq k}$ to denote $\reachone_{\alpha,\geq k}$ for some reaction $\alpha$ of $\calN$ and we define the transitive and reflexive closure of $\reachone_{\calN,\geq k}$ by $\reach_{\calN,\geq k}$. Moreover, we define $\pre_{\calN,\geq k}(\vc) = \{ \vc' \in \N^\Lambda \mid \vc' \reach_{\calN,\geq k} \vc \}$. As usual we omit the subscript $\calN$, and write simply $\pre_{\geq k}$ when the CRN is clear from the context.

Recall that for stably deciding CRDs it holds that for all input states $\vc \in \N^\Sigma \setminus \{\vec{0}\}$, we have $\post(\imath(\vc)) \subseteq \pre(\mathcal{S}_b)$ for some $b \in \{0,1\}$. In other words, from every state reachable from an input state, we can reach an output-$b$ stable state. 

We now define when such a CRD is speed-fault free \cite{SpeedFaults/DC/CCDS/2017}.

\begin{definition}\label{def:speed_fault_free}
We say that a stably deciding CRD with only uni- and bimolecular reactions is \emph{speed-fault free} if there is a distinguished input species $F \in \Sigma$ such that for all $k \in \N$ there is an $n \in \N$ such that for all input states $\vc \in \N^\Sigma \setminus \{\vec{0}\}$ with $\#_\vc F \geq n$, $\post(\imath(\vc)) \subseteq \pre_{\geq k}(\mathcal{S}_b)$ for some $b \in \{0,1\}$.
\end{definition}

The distinguished input species $F$ of Definition~\ref{def:speed_fault_free} is called the \emph{fuel} species. Definition~\ref{def:speed_fault_free} says that a stably deciding CRD is speed-fault free if any state $\vc'$ reachable from some input state having at least $n$ fuel molecules can reach an output-stable state using only $k$-fast reactions.

The next example is essentially taken from \cite{SpeedFaults/DC/CCDS/2017}.
\begin{example}\label{ex:CRD_existence_check}
Consider the problem of deciding whether or not there is at least one $A_1$ or $A_2$ molecule and no $A_3$ molecule. Let $\Sigma = \{ A_1, A_2, A_3, F \}$, where $F$ is the fuel species. Identify the species $A_1$, $A_2$, $A_3$, $F$ with $X_{100}$, $X_{010}$, $X_{001}$, $X_{000}$, respectively. These subscripts are bit-vectors identifying the presence or absence of the $A_i$ molecules. We introduce species $X_{b_1 b_2 b_3}$ for every bit-vector $b_1 b_2 b_3$ and we introduce bimolecular reactions $X_{v} + X_w \to 2X_{\mathrm{OR}(v,w)}$ where $v \neq w$ and $\mathrm{OR}$ denotes bitwise OR. For example, we introduce the reactions $X_{000} + X_{001} \to 2X_{001}$ and $X_{110} + X_{011} \to 2X_{111}$. One can easily verify that the corresponding CRN $\calN$ eventually halts where all molecules are of the same species $X_{b_1 b_2 b_3}$ where the $b_i$'s indicate the presence ($b_i = 1$) or absence ($b_i = 0$) of $A_i$-molecules in the input state. The species $X_{b_1 b_2 b_3}$ with $(b_1 = 1 \lor b_2 = 1) \land b_3 = 0$ precisely satisfy the above given predicate and we define these to be exactly the yes-voters. The total number of molecules in a state does not change when reactions take place and so a halting state is nonzero if and only if the input state is nonzero. Hence the obtained CRD $\calD$ haltingly (and, therefore, stably) decides the given predicate. Note that the CRD also works fine if we omit the fuel species $F$ and the reactions in which $F$ appears. The sole purpose of species $F$ is to make $\calD$ speed-fault free (and, in this way, the ability to ``boost'' the computation by increasing the number of $F$-molecules). It is easy to see that $\calD$ is speed-fault free. Indeed, note that there are $2^3$ species $X_{b_1 b_2 b_3}$. Thus at least one species $X_{b_1 b_2 b_3}$ with at least $t/2^3$ molecules, where $t$ be the total number of molecules (recall that $t$ does not change when reactions take place). For any $k \in \N$, take $n := k \cdot 2^3$. Then $t/2^3 \geq n/2^3 = k$, so for any non-halting state $\vc$ there is a $k$-fast reaction for $\vc$, and thus $\calD$ is speed-fault free.
\end{example}

The following has been shown in \cite{SpeedFaults/DC/CCDS/2017}.
\begin{theorem}[\cite{SpeedFaults/DC/CCDS/2017}]\label{thm:speed_fault_free}
Let $X \subseteq \N^{\Sigma}$ for some finite set $\Sigma$. Then $X$ is obtained by finite unions, intersections, and complementations of sets of the form $S_A = \{ \vc \in \N^{\Sigma} \mid \#_\vc A = 0 \}$ with $A \in \Sigma$ if and only if there is a speed-fault-free stably deciding CRD $\calD$ with only uni- and bimolecular reactions that recognizes a set $X' \subseteq \N^{\Sigma'}$ with $X'|_{\Sigma} = X$ and $\Sigma' = \Sigma \cup \{F\}$ where $F$ is the fuel species of $\calD$.
\end{theorem}
So, essentially, such CRDs can only distinguish between existence and non-existence of molecules of the input species. Consequently, they can decide predicates of the form ``there is at least one molecule of $A$'' but \emph{not} predicates of the form ``there are at least two molecules of $A$''. We note that speed-fault freeness is only an indication (no guarantee) for fast computation. However, the approach described in Example~\ref{ex:CRD_existence_check} to determine (non-)existence of molecules in the input state by speed-fault free CRDs can be shown to fast compute assuming the standard stochastic CRN model (cf.\ Subsection~\ref{ssec:stoch_CRNs}), for details see \cite{SpeedFaults/DC/CCDS/2017}.

\begin{remark}
Non-speed-fault-free stably deciding CRDs $\calD$ are not necessarily slower, assuming the standard stochastic CRN model, than speed-fault-free stably deciding CRDs. Indeed, the existence of a single $\vc'$ reachable from some input state $\vc$ that does \emph{not} have a $k$-fast trajectory to an output stable state (i.e., $\vc' \in \pre(\mathcal{S}_b)$, but $\vc' \notin \pre_{\geq k}(\mathcal{S}_b)$) may have little effect on the speed of the CRD if $\vc'$ is unlikely to be reached from the input state $\vc$ in the first place. 
\end{remark}

The notions and results concerning speed faults turned out to be a stepping stone to prove time lower bounds for problems like leader election (cf.\ Subsection~\ref{ssec:leader_election}) in the standard stochastic CRN model.

\begin{remark}
In Definition~\ref{def:speed_fault_free}, the fuel species $F$ is defined to be in the input alphabet $\Sigma$. However, in \cite{SpeedFaults/DC/CCDS/2017} the fuel species is not defined to be in $\Sigma$. Consequently, the formulation of Theorem~\ref{thm:speed_fault_free} above is slightly different than its corresponding formulation in \cite{SpeedFaults/DC/CCDS/2017}. The formulation of Theorem~\ref{thm:speed_fault_free} above raises the question whether the whole set $X'$ (not merely $X'|_\Sigma = X$) is obtained by finite unions, intersections, and complementations of sets of the form $S_A$'s with $A \in \Sigma'$. Even more generally, one can also consider a definition of speed-fault freeness without explicit fuel species $F$, where the condition $\#_\vc F \geq n$ is replaced by $\| \vc \| \geq n$.
\end{remark}

\begin{remark}
We remark that Theorem~\ref{thm:speed_fault_free} is shown in \cite{SpeedFaults/DC/CCDS/2017} for the more general class of CRDs which have an initial ``context'' $\vc \in \N^{\Lambda \setminus \Sigma}$ (in fact, this complicates the proof of Theorem~\ref{thm:speed_fault_free} significantly). This initial context is present at the start of the computation along with the input (see Subsection~\ref{ssec:leader_election} for a definition). The notion of a CRD as defined above corresponds to the case where the initial context $\vc$ is $\vec{0}$. 
\end{remark}

\section{Computing with stochastic chemical reaction networks}\label{sec:comp_stoch_CRNs}
The computational CRN models of Section~\ref{sec:comp_CRNs} only concern reachability of states, and so their results are independent of stochastics (i.e., how likely a certain state is reached). In this section we recall the well-known standard stochastic model for CRNs and then show, assuming this stochastic model, that CRNs can perform Turing-universal computation if we allow an arbitrary small error probability. We also illustrate the computational mechanism of stochastic CRNs by considering the leader election problem and computing probability distributions.

\subsection{Stochastic chemical reaction networks}\label{ssec:stoch_CRNs}
We first recall the (standard) stochastic model for CRNs \cite{mcquarrie_1967/StochCRN}. In this section we use several notions and notation from \cite{DBLP:journals/nc/SoloveichikCWB08}.

In the stochastic model for CRNs, each reaction $\alpha$ has a value $k_\alpha \in \R^{+}$ called the \emph{rate constant}. The \emph{volume} $v \in \R_{>0}$ represents the volume of the well-mixed solution, and as such it determines the expected time for two fixed molecules in the well-mixed solution to meet. The larger the volume, the slower reactions with more than one reactant will take place. Due to physical constraints, the ratio $\|\vc\|/v$ is bounded above for well-mixed solutions --- this is called the \emph{finite density constraint}. Consequently, one cannot make the volume arbitrarily small, and the cardinality of a state can only grow unboundedly when the volume grows unboundedly too by continuously diluting the solution.

Let $v \in \N$ be a fixed volume. Define the \emph{propensity} $\rho(\vc,\alpha)$ of a reaction $\alpha  = (\vr,\vp)$ in a state $\vc$ as
\[
\frac{k_\alpha}{v^{\|\vr\|-1}} \prod_{X \in S} \vc(X)(\vc(X)-1) \cdots (\vc(X)-(\vr(X)-1)),
\]
see, e.g., \cite{AndersonKurtz2011}. Here, the product counts the number of ways one can pick all the molecules of $\vr$ from the state $\vc$, the fraction $\frac{1}{v^{\|\vr\|-1}}$ represents the likelihood that all molecules of $\vr$ simultaneously meet, and $k_\alpha$ represents the likelihood that when these molecules meet, they will react (i.e., the reaction will take place).

In particular, if $\alpha$ is unimolecular, then $\rho(\vc,\alpha) = k_\alpha \vc(X)$ where $X \in \Lambda$ such that $\vr(X) = 1$. If $\alpha$ is bimolecular, then $\rho(\vc,\alpha)$ is equal to $\frac{k_\alpha}{v} \vc(X_1) \vc(X_2)$ in the case where $X_1, X_2 \in \Lambda$ are distinct such that $\vr(X_1) = \vr(X_2) = 1$, and is equal to $\frac{k_\alpha}{v} \vc(X) (\vc(X)-1)$ in the case where $\vr(X) = 2$.

For $\vc, \vc' \in \N^\Lambda$, define the \emph{transition rate} from $\vc$ to $\vc'$ as 
\[
\rho(\vc,\vc') := \sum_{\alpha \in R(\calN), \vc \reachone_\alpha \vc'} \rho(\vc,\alpha).
\]
We remark that the transition rates define a continuous-time Markov chain on the set of states of $\calN$. However, in this paper we assume no familiarity with Markov chains.

The duration for some reaction to take place within the state $\vc$ is an exponential random variable, i.e., a continuous random variable that depends only on the current state and not on the amount of time elapsed (the random variable is ``memoryless''), with rate 
\[
\rho(\vc,\calN) := \sum_{\alpha \in R(\calN)} \rho(\vc,\alpha).
\]
The probability that $\alpha \in R(\calN)$ is the next reaction to occur in $\vc$ is equal to $\rho(\vc,\alpha)/\rho(\vc,\calN)$. In particular, the expected time for some reaction to take place in $\vc$ is $1/\rho(\vc,\calN)$ (if $\rho(\vc,\calN)=0$, then no reaction can occur in $\vc$). 

Coming back to the notion of speed faults (cf.\ Subsection~\ref{ssec:speed_faults}), we have, in particular, that increasing the molecular count of one of the reactant species of a reaction $\alpha$, increases its propensity, and therefore decreases the expected time of this reaction to take place (assuming $\alpha$ is the next reaction to take place). In other words, the reaction will indeed be faster.

\begin{example}
Consider a CRN with the reactions $\alpha = A + B \to C$ and $\beta = 2A \to C$. For any state $\vc$, we have $\rho(\vc,\alpha) = \frac{k_\alpha}{v} \vc(A) \vc(B)$ and $\rho(\vc,\beta) = \frac{k_\beta}{v} \vc(A) (\vc(A)-1)$. Since the expected time for some reaction to take place in $\vc$ is $1/\rho(\vc,\calN)$, increasing the number of molecules of $A$ and $B$ will decrease this expected time. If both reactions are applicable to $\vc$ (i.e., $\vc$ has at least 2 molecules of $A$ and at least 1 molecule of $B$), then the probability that $\alpha$ is the next reaction to occur in $\vc$ is 
\[
\frac{\rho(\vc,\alpha)}{\rho(\vc,\alpha)+\rho(\vc,\beta)} = \frac{1}{1+\frac{k_\beta}{k_\alpha} \cdot \frac{c(A)-1}{c(B)}},
\]
which is tending to $1$ by increasing the number of $B$-molecules compared to $A$-molecules.
\end{example}

Note that because of the finite density constraint, one cannot arbitrarily speed up the computation by decreasing $v$. Similarly, one cannot arbitrarily increase $\|\vc\|$ without increasing $v$.

Rate constants of chemical reactions are very difficult to control because they depend on the molecular structure of their reactants. Therefore, computational CRN models are often designed to work for any choice of rate constants. That is, we assume we cannot set the rate constants ourselves and so, e.g., the rate constants appear as undetermined constants in various results, such as time complexity results. For notational convenience, we assume in this paper that all rate constants are equal to some fixed value $k$. Also, for notational convenience, by a ``stochastic CRN'' we mean a CRN with rate constants for each reaction that operates in the above described way.

\subsection{Turing-universal computation by stochastic chemical reaction networks with possible errors}\label{ssec:SCRN_Turing_universal}
In this subsection we show that stochastic CRNs can simulate any Turing machine if we allow an arbitrary small nonzero probability of error. Various computational models are Turing universal, and here we follow \cite{DBLP:journals/nc/SoloveichikCWB08} by simulating deterministic counter automata (which are Turing universal \cite{Minsky/RM/1961,HopUll/IntroFormalLang}) by stochastic CRNs. See also \cite{DBLP:journals/nc/SoloveichikCWB08} for a direct simulation of Turing machines by stochastic CRNs.

We briefly recall the notion of a counter automaton (also sometimes called register machine in the literature), see, e.g., \cite{Minsky/RM/1961,HopUll/IntroFormalLang} for a more elaborate treatment. A (deterministic) \emph{counter automaton} $M$ is a finite state automaton, with distinguished start and halting states $q_{\mathrm{start}}$ and $q_{\mathrm{halt}}$, augmented with a finite number of \emph{counters} (also called registers in the literature) which can each hold an arbitrary non-negative integer. For any state $q$ of $M$, there is an instruction
\begin{itemize}
\item $\mathrm{inc}(q,c,q')$ which increments counter $c$ by $1$ and then moves to state $q'$ or
\item $\mathrm{dec}(q,c,q',q'')$ which either (1) decrements counter $c$ by $1$ and moves to state $q'$ if the value of $c$ is nonzero or (2) moves to state $q''$ (leaving the value of $c$ unchanged) if the value of $c$ is zero. 
\end{itemize}
We assume that for each state $q$ there is exactly one such instruction (hence the adjective ``deterministic''). The input of a counter automaton is a nonnegative integer that is stored in the input counter (a distinguished counter) and the input is accepted when, starting in the start state, the computation eventually reaches the halting state.

Given a counter automaton $M$ we define a CRN $\calN_M$ simulating $M$ with low probability of error as follows. The set $\Lambda$ of species of $\calN_M$ is equal to $Q \cup C$, where $Q$ is the (finite) set of states and $C$ is the (finite) set of counters of $M$ (we assume without loss of generality that $Q$ and $C$ are disjoint). Furthermore, for each instruction $\mathrm{inc}(q,c,q')$ we introduce the reaction $q \to c + q'$ in $\calN_M$ and for each instruction $\mathrm{dec}(q,c,q',q'')$ we introduce two reactions $q + c \to q'$ and $q \to q''$. Let $\imath$ be the input counter of $M$ and $\nu \in \N$ be an input value of $M$. Then we take as the input state $\vi_\nu$ of $\calN_M$ the state with one molecule of the start state $q_{\mathrm{start}}$ of $M$ and $\nu$ molecules of species $\imath$.

The idea is that during the computation there is exactly one molecule of a species in $Q$, which represents the current state of $M$, and, for each $c \in C$, the number of $c$-molecules is equal to the value of the counter $c$ in $M$. Note that the reaction $q \to c + q'$ correctly simulates the instruction $\mathrm{inc}(q,c,q')$. Moreover, if the value of counter $c$ is zero, then $\mathrm{dec}(q,c,q',q'')$ is correctly simulated by $q \to q''$ (and $q + c \to q'$ cannot take place). If the value of counter $c$ is nonzero, then $\mathrm{dec}(q,c,q',q'')$ is correctly simulated by $q + c \to q'$, however reaction $q \to q''$ can also take place. In the latter case, i.e., when reaction $q \to q''$ takes place with $c$-molecules present, the computation is in error. To make the chance of error arbitrary small, we modify the reaction $q \to q''$ to make it arbitrary slow: indeed, the slower this reaction, the more likely the correct reaction $q + c \to q'$ is taken instead. In this way, we trade computation speed for a lower probability of error. To accomplish this trade, the reaction $q \to q''$ is replaced by the following reactions: $T_i + D \to T_{i+1} + D$ and $T_{i+1} \to T_i$ for $i \in \{1, \ldots, l-1\}$ and some nonnegative integer $l$, and $T_1 + q \to q'' + T_l$. Here, $T_1, \ldots, T_l$, and $D$ are all new species. Moreover, $q + c \to q'$ is replaced by the reaction $q + c \to q' + D$. Also, the input state $\vi_\nu$ of $\calN_M$ now also contains one $T_l$-molecule and a sufficiently large number of $D$-molecules (depending on the rate constants and volume). The higher the number of $D$-molecules, the longer it takes for a $T_l$ molecule to convert to a $T_1$ molecule, while in turn a $T_1$ molecule is required for the reaction $T_1 + q \to q'' + T_l$ to take place (which corresponds to moving from state $q$ to state $q''$). In fact, the production of $T_1$ takes more and more time as the computation of $M$ progresses since each transition from $q$ to $q'$ by decreasing counter $c$ introduces a new $D$-molecule. Since the value $l$ is not fixed, we denote the resulting CRN by $\calN_{M,l}$.

By \cite[Theorem~3.1 and Section 4]{DBLP:journals/nc/SoloveichikCWB08} we have the following.
\begin{theorem}[\cite{DBLP:journals/nc/SoloveichikCWB08}]\label{thm:SCRN_universal_TM}
Let $M$ be a counter automaton, $\delta > 0$, and $\nu \in \N$. Then there is an $l \in \N$, such that $\calN_{M,l}$ on input $\vi_\nu$ simulates $M$ with a cumulative error probability of at most $\delta$.
\end{theorem}

See \cite{DBLP:journals/nc/SoloveichikCWB08} for upper bounds on the expected computation time and for a faster computation by simulating Turing machines instead of counter automata. Finally, we remark that in \cite{DotySoloveichik/ZeroError} an analog of Theorem~\ref{thm:SCRN_universal_TM} is obtained with error probability zero in terms of ``limit-stable'' computations: although there might be (infinite) trajectories that lead to an error, these ``wrong'' trajectories together do not contribute to a positive error probability. While the notions of error probability zero and error-free coincide when each state has only a finite number of reachable states, these notions diverge when states can have an infinite number of reachable states.

We also mention that, independently, a similar approach of simulating Turing machines (via counter automata with multiplication and division) was taken in \cite{AngluinAE08a/DISTJournal/FastWithLeader} in the context of population protocols (see Subsection~\ref{ssec:PopProt} for a discussion on the relation between CRNs and population protocols). Finally, we mention that finite circuit computation was shown to be achievable using CRNs in \cite{PhysRevLett/CRNSimFiniteCircuits} (despite its title, the paper does not show Turing universality).

\subsection{Leader election}\label{ssec:leader_election}
We now turn to the problem of \emph{leader election}. To motivate this problem, consider a natural extension of the notion of a CRD $\calD = (\calN,\Sigma,\Lambda_0,\Lambda_1)$, where we extend $\calD$ by a vector $\jmath \in \N^{\Lambda \setminus \Sigma}$, called the \emph{context}, to obtain the 5-tuple $\calD' = (\calN,\Sigma,\Lambda_0,\Lambda_1,\jmath)$. The molecules of $\jmath$ are assumed to be present at the start of a computation. Hence, the initial state consists of the input molecules and the molecules of the context. It turns out that the notion of a CRD with context does not lead to a (significant) increase in computational power, i.e., CRDs with context can also only compute semilinear sets \cite{DBLP:journals/dc/AngluinAER07} (the only difference is that CRDs with context can also compute the semilinear sets $X$ with $\vec{0} \in X$).

While the expressive power of the class of CRDs with context is equal to that of the class of ordinary CRDs, it is natural to wonder whether or not there are predicates that can compute \emph{faster} using CRDs with context compared to ordinary CRDs.

An interesting special case of this problem is where $\|\jmath\|=1$, the (unique) molecule of $\jmath$ is called the \emph{leader} of $\calD'$. For designing a CRN that computes a given predicate, it is often convenient to have a leader. Intuitively, a leader can ``guide'' the computation much like the control flow dictates the computation for an ordinary computer program. Indeed, in \cite{AngluinAE08a/DISTJournal/FastWithLeader} it has been shown that various predicates can be computed efficiently if a leader is present. Conversely, various other predicates have been shown to be slow without a leader \cite{BellevilleDS17/ICALP/LeaderlessPopProt}. 

In the absence of a leader, one can construct a leader (i.e., a single molecule of some given species) --- this is called \emph{leader election}. It is straightforward to elect a leader as follows: assuming there is at least one molecule of $L \in \Sigma$, then the reaction $L + L \to L$ eventually results in a single $L$-molecule. However, the process of constructing a leader in this way is slow: $O(n)$ expected time with $n$ molecules of $L$ present in the initial state. Indeed, leader election turns out to be necessarily slow \cite{DotyS/DISTJournal/LeaderElectionLinTime}.

\subsection{Computing probability distributions}\label{ssec:comp_prob_dist}

A different way to define the computation of a stochastic CRN is through probability distributions \cite{FettBR/CRN,CardelliKL16/DNA22/ProgProbDistr}. 

Given a stochastic CRN $\calN$ and a state $\vc$, we denote by $\mathrm{Prob}_{\calN,\vc}(t,\vd)$, for $t \in \R_{\geq 0}$ and state $\vd$, the probability of reaching state $\vd$ at time $t$. Note that for fixed $t_1 \in \R_{\geq 0}$, $\mathrm{Prob}_{\calN,\vc}(t_1,\vd)$ can be seen as a function sending states $\vd \in \N^\Lambda$ to values in the real interval $[0,1]$. Also note that $\sum_{\vd \in \N^\Lambda} \mathrm{Prob}_{\calN,\vc}(t_1,\vd)$ exists and is equal to $1$. We call such functions $f: \N^\Lambda \to [0,1]$ with $\sum_{\vd \in \N^\Lambda} f(\vd) = 1$ \emph{probability mass functions}.

Similarly, if $\pi_{\calN,\vc}(\vd) := \lim_{t \to \infty} \mathrm{Prob}_{\calN,\vc}(t,\vd)$ exists, then $\pi_{\calN,\vc}(\vd)$ is a probability mass function. Intuitively, $\pi_{\calN,\vc}(\vd)$ describes the long-term probability distribution of the states of $\calN$ starting from $\vc$.

For probability mass functions $f_1$ and $f_2$, we define $d(f_1,f_2) = \sum_{\vd \in \N^\Lambda} |f_1(\vd) - f_2(\vd)|$. Note that $d(f_1,f_2)$ is well defined (in fact, $d(f_1,f_2) \leq 2$). The \emph{support} of a probability mass function $f$ is the set of states $\vc$ such that $f(\vc)$ is nonzero.

The next result shows that arbitrary probability mass functions can be approximated by stochastic CRNs.
\begin{theorem}[\cite{CardelliKL16/DNA22/ProgProbDistr}]\label{thm:pmf_CRN}
Let $f: \N^\Lambda \to [0,1]$ be a probability mass function and $\epsilon > 0$. Then there exists a stochastic CRN $\calN$ and a state $\vc$ such that $\pi_{\calN,\vc}$ exists and $d(f,\pi_{\calN,\vc}) < \epsilon$. If $f$ has moreover finite support, then there exists a stochastic CRN $\calN$ and a state $\vc$ such that $f=\pi_{\calN,\vc}$.
\end{theorem}
In the case where $f=\pi_{\calN,\vc}$, we say that $\calN$ \emph{computes} $f$ starting in $\vc$. The proof of Theorem~\ref{thm:pmf_CRN} first shows the exact computation result (the case where $f$ has finite support) and then observes that the approximate computation result holds since the probability mass functions with finite support are dense for all probability mass functions with countable domain $\N^\Lambda$ under the distance metric $d$. It is an open question to characterize the set of probability mass functions that can be (exactly) computed (of course, this set includes all probability mass functions with finite support).

Furthermore, in \cite{CardelliKL16/DNA22/ProgProbDistr} a \emph{calculus} for probability mass functions that are zero for all but a finite number of states is defined such that any such probability mass function can be obtained from a formula in this calculus. The operators have been shown to be implementable using CRNs \cite{CardelliKL16/DNA22/ProgProbDistr}. In this way, a \emph{programming language} for probability mass functions based on CRNs is obtained.

\section{Computing with continuous chemical reaction networks}\label{sec:CompContinuousCRNs}

\subsection{Continuous chemical reaction networks}\label{ssec:ContinuousCRNs}
Until now, a state of a CRN is a vector describing the molecular \emph{counts} $\#X$ of the species $X$. Such a state is also called a \emph{discrete state}. The larger these molecular counts, the more the stochastic model tends to the continuous mass-action kinetics model, which we call simply the continuous CRN model in this paper, up to some point in time \cite{Kurtz/ContLimitDiscrete}. We remark however that this point in time where divergence of the continuous mass-action kinetics model with the stochastic model can happen is rather soon, namely logarithmic in the number of molecules.

Denote by $\R_{\geq 0}$ the set of nonnegative real numbers. In the continuous CRN model, a state is a $\R_{\geq 0}$-valued vector describing the molecular \emph{concentrations} $[X]$ of the species $X$. To distinguish both types of states, we call a state describing molecular concentrations, a \emph{continuous state}. For notational convenience, by a \emph{discrete CRN} (\emph{continuous CRN}, resp.) we mean a CRN that uses discrete (continuous, resp.) states. Note that a stochastic CRN is a particular kind of discrete CRN.

A continuous state evolves continuously (with $\R_{\geq 0}$-valued time variable $t$) according to a set of ordinary differential equations (ODEs). To define these ODEs, we first recall the notion of a \emph{stoichiometry matrix} $M$ of $\calN$. The rows and columns of $M$ corresponds to the species $X$ and reactions $\alpha$ of $\calN$, respectively, and each entry $M_{X,\alpha}$ describes the net change of the $X$-molecules when reaction $\alpha$ takes place. For example, consider the CRN $\calN$ with reactions $\alpha = A+B \to C$ and $\beta = 2C + B \to 2A + B$. Then the stoichiometry matrix of $\calN$ is as follows
\[
M = 
\bordermatrix{
~ & \alpha & \beta \cr
A & -1 & 2 \cr
B & -1 & 0 \cr
C &  1 & -2 \cr
}.
\]
The concentration of some species $X$ changes according to the ODE
\[
\dv{[X]}{t} = \sum_{\alpha = (\vr,\vp) \in R} k_\alpha M_{X,\alpha} \prod_{Y \in \Lambda} [Y]^{\vr(Y)},
\]
where $k_\alpha$ denotes the rate constant of reaction $\alpha$. Thus each reaction $\alpha$ contributes to a change in concentration of species $X$ that is equal to the product of the reactant concentrations of $\alpha$, its rate constant $k_\alpha$, and the difference of the number of times $X$ is a product of $\alpha$ minus the number of times $X$ is a reactant of $\alpha$ (so, e.g., the contribution of $\alpha$ to the concentration of $X$ is negative when $X$ appears as a reactant, but not as a product of $\alpha$).

So, in the given example, a continuous state changes according to the following set of ordinary differential equations:
\begin{align*}
\dv{[A]}{t} &= -k_\alpha[A][B] + 2k_\beta[C]^2[B] \\
\dv{[B]}{t} &= -k_\alpha[A][B] \\
\dv{[C]}{t} &= k_\alpha[A][B] - 2k_\beta[C]^2[B]
\end{align*}
where $k_\alpha$ and $k_\beta$ are the rate constants of $\alpha$ and $\beta$, respectively. We remark that, since states change deterministically in the continuous CRN model, this model is often called the ``deterministic'' CRN model --- however, we do not use this terminology here to avoid possible confusion with the computational CRN models of Section~\ref{sec:comp_CRNs} that also have various deterministic aspects.

We now briefly sketch the computational model of continuous CRNs from \cite{DBLP:conf/cmsb/FagesGBP17}, see that reference for the (involved) formal definition.
Roughly speaking, a function $f: \R_{\geq 0} \to \R_{\geq 0}$ is called \emph{chemically-computable} if there exists a continuous CRN $\mathcal{N}$ and a $\Lambda$-indexed vector $q(x)$, where each entry of $q(x)$ is a polynomial in variable $x$ with coefficients from $\R_{\geq 0}$, such that, for all $z \in \R_{\geq 0}$, starting in state $q(z)$, the state $\vec{c}$ of the CRN evolves in such a way that $\vec{c}(S)$, for some distinguished species $S$, approaches the value $f(z)$ as $t \to \infty$. In other words, to compute $f(z)$, $q$ maps $z$ to the initial state of the CRN and the value $f(z)$ is represented by a distinguished entry of the state to which the CRN converges. By using a ``dual rail'' approach that is similar to the one discussion in Subsection~\ref{ssec:rateIndepCRNcomp} below, one can extend the notion of chemically-computable to functions $f: \R \to \R$, i.e., where the domain and codomain is $\R$. 

It is then shown in \cite{DBLP:conf/cmsb/FagesGBP17} that chemically-computable functions are exactly the functions computable by so-called General Purpose Analog Computers as defined in \cite{Bournez/2007/AnalogComp} (which is somewhat different from the original definition in \cite{Shannon/AnalogComp}). In turn, General Purpose Analog Computers (as defined in \cite{Bournez/2007/AnalogComp}) are computationally equivalent to Turing machines. In this way, this computational model of continuous CRNs is Turing universal.

\subsection{Rate-independent computation with continuous chemical reaction networks}\label{ssec:rateIndepCRNcomp}

Early work on the computational power of continuous CRNs includes \cite{CRN/MassActionComp/2009/Buisman}, where it is shown that various numerical operations such as addition and multiplication can be implemented by continuous CRNs assuming the rate constants of the used reactions can be tuned. Since rate constants are however notoriously difficult to tune, a computational model for continuous CRNs has been introduced in \cite{DBLP:conf/innovations/ChenDS14} that works independently of the rate constants of the individual reactions. In this subsection we discuss the computational model of \cite{DBLP:conf/innovations/ChenDS14}. We remark that another rate-independent model of computation for continuous CRNs has been studied in \cite{Riedel_rate_indep_CCRNs}.

The computational model for continuous CRNs in \cite{DBLP:conf/innovations/ChenDS14} is an analog of the computational model in Section~\ref{sec:comp_CRNs} but with a reachability function that deals with continuous states instead of (integer-valued) states. 

We say that a reaction $\alpha = (\vr,\vp)$ is \emph{applicable} to a continuous state $\vc$ if for all species $X$, $\vr(X) > 0$ implies that $\vc(X) > 0$. For continuous states $\vc$ and $\vd$ and $\vec{u} \in \R_{\geq 0}^{R}$, we write $\vc \reachone_\vec{u} \vd$ if $\vc + M \vec{u} = \vd$, where $M$ is the stoichiometry matrix, and $\vec{u}(\alpha) > 0$ implies that $\alpha$ is applicable to $\vc$. Here $\vec{u}(\alpha) \in \R_{\geq 0}$ represents the ``amount'' of reaction $\alpha$ to occur and so $(M \vec{u})(X)$ represents the change in concentration of $X$ when all reactions take place in the amounts described by $\vec{u}$. Therefore, $\vd = \vc + M \vec{u}$ is the state obtained from state $\vc$ when the reactions take place according to $\vec{u}$.

We say that $\vd$ is \emph{straight-line reachable} from $\vc$, denoted by $\vc \reachone \vd$, if there is a $\vec{u} \in \R_{\geq 0}^{R}$ such that $\vc \reachone_\vec{u} \vd$. As usual, the transitive and reflexive closure of $\reachone$ is denoted by $\reach$. We say that $\vd$ is \emph{segment-reachable} from $\vc$ if $\vc \reach \vd$. Note that segment-reachability is quite different from the reachability notion implied by the ODEs of Subsection~\ref{ssec:ContinuousCRNs} (which is very much rate dependent). Indeed, Subsection~\ref{ssec:ContinuousCRNs} implies a definition of reachability such that a continuous state $\vd$ is reachable from $\vc$ if $\vd$ corresponds to the continuous state at time $t>0$ starting from continuous state $\vc$ at time $t=0$. While the two notions are quite different, \cite{DBLP:conf/innovations/ChenDS14} shows some relationships between these two notions of reachability. In particular, if a state $\vd$ is mass-action reachable from state $\vc$, then $\vd$ is segment-reachable from $\vc$.

With the notion of reachability defined in this subsection in place, one can straightforwardly define the continuous analogs of stably deciding for chemical reaction deciders (CRDs) and stably computing for chemical reaction computers (CRCs) of Section~\ref{sec:comp_CRNs}, see \cite{DBLP:conf/innovations/ChenDS14}. Let us call the continuous analog of stably computing, \emph{$\R_{\geq 0}$-stably computing}.

\begin{example}\label{ex:minmaxCRC_cont}
One verifies that the CRCs described in Examples~\ref{ex:minCRC} and \ref{ex:maxCRC} $\R$-stably compute the $\min(x,y)$ and $\max(x,y)$ functions where $x$ and $y$ are, more generally, in $\R_{\geq 0}$ instead of in $\N$.
\end{example}

In order to compute general real-valued functions (which allow negative values), we additionally need the notion of a ``dual-rail'' representation. A \emph{dual-rail representation} of $f: \R^\Sigma \to \R^\Gamma$ is a function $\hat f: \R_{\geq 0}^\Sigma \times \R_{\geq 0}^\Sigma \to \R_{\geq 0}^\Gamma \times \R_{\geq 0}^\Gamma$ such that for all $\vx^+,\vx^- \in \R_{\geq 0}^\Sigma$ and $\vy^+,\vy^- \in \R_{\geq 0}^\Gamma$, $\hat f(\vx^+,\vx^-) = (\vy^+,\vy^-)$ implies that $f(\vx^+ - \vx^-) = \vy^+ - \vy^-$. We remark here that CRNs that compute using dual-rail representations of functions can be straightforwardly composed in contrast to CRNs that compute functions in the ordinary way --- this is true for both discrete and continuous CRNs.

The following example is taken from \cite{DBLP:conf/innovations/ChenDS14}.
\begin{example}
Consider the $\min(x,y)$ and $\max(x,y)$ functions over $\R$, i.e., $\min$ and $\max$ compute the minimum and maximum of two real numbers $x$ and $y$.

Let $\Sigma = \{X,Y\}$, $\Gamma = \{Z\}$, $\hat\Sigma = \{X^+,X^-,Y^+,Y^-\}$, and $\hat\Gamma = \{Z^+,Z^-\}$. A dual-rail representation $\widehat \min: \R^{\hat \Sigma} \to \R^{\hat \Gamma}$ of $\min: \R^\Sigma \to \R^\Gamma$ can be computed by the reactions
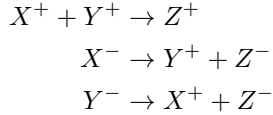
\begin{align*}
X^+ + Y^+ &\to Z^+ \\
X^- &\to Y^+ + Z^- \\
Y^- &\to X^+ + Z^-
\end{align*}
To see this, first notice that both the values $(\# X^+ - \# X^-) + (\# Z^+ - \# Z^-)$ and $(\# Y^+ - \# Y^-) + (\# Z^+ - \# Z^-)$ are invariant under applying these reactions. Also notice that for any state a halting state is reachable: the last two reactions can take place until no $X^-$ and $Y^-$ molecules are present and then the first reaction can take place until the $X^+$ or $Y^+$ molecules are exhausted.

Let $\vi$ be an initial state, i.e., consisting of only $X^+$, $X^-$, $Y^+$, and $Y^-$ molecules. It is easy to see that the CRN has halted in some state $\vc$ precisely when $\#_\vc X^- = \#_\vc Y^- = 0$ and either $\#_\vc X^+ = 0$ or $\#_\vc Y^+ = 0$. By the invariance properties $\#_\vi X^+ - \#_\vi X^- = (\#_\vi X^+ - \#_\vi X^-) + (\#_\vi Z^+ - \#_\vi Z^-) = (\#_\vc X^+ - \#_\vc X^-) + (\#_\vc Z^+ - \#_\vc Z^-) = \#_\vc X^+ + (\#_\vc Z^+ - \#_\vc Z^-)$ and similarly for $\#_\vi Y^+ - \#_\vi Y^-$. In the case where $\#_\vc X^+ = 0$, we have that $\#_\vi Y^+ - \#_\vi Y^- \geq \#_\vi X^+ - \#_\vi X^- = \#_\vc Z^+ - \#_\vc Z^-$ and so $\#_\vc Z^+ - \#_\vc Z^-$ is indeed equal to $\min(\#_\vi X^+ - \#_\vi X^-,\#_\vi Y^+ - \#_\vi Y^-)$. The case where $\#_\vc Y^+ = 0$ is analogous, and so we conclude that this CRN (more precisely, CRC) indeed computes $\widehat \min$.

Note that the special case where $\#_{\vi}X^- = \#_{\vi}Y^- = 0$ essentially corresponds to the usual ``single-rail'' computation of $\min$, cf.\ Example~\ref{ex:minmaxCRC_cont}, since then only the reaction $X^+ + Y^+ \to Z^+$ can take place.

Since $\max(x,y) = -\min(-x,-y)$, a dual-rail representation $\widehat \max: \R^{\hat \Sigma} \to \R^{\hat \Gamma}$ of $\max: \R^\Sigma \to \R^\Gamma$ is obtained from $\widehat \min$ by reversing the roles of the ``plus'' and ``minus'' species (i.e., flipping the superscript). Thus, $\widehat \max$ can be computed by the reactions
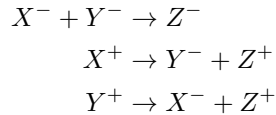
\begin{align*}
X^- + Y^- &\to Z^- \\
X^+ &\to Y^- + Z^+ \\
Y^+ &\to X^- + Z^+
\end{align*}
\end{example}

Let $f: \R^\Sigma \to \R$ be a function. Then $f$ is called \emph{rational linear} if there is a $\vec{a} \in \mathbb{Q}^\Sigma$ such that $f(\vx) = \vec{a}\cdot\vx$, where $\cdot$ denotes the dot product. Moreover, $f$ is called \emph{piecewise rational linear} if there is a finite set $S$ of rational linear functions such that for every $\vx \in \R^\Sigma$, $f(\vx) = g(\vx)$ for some $g \in S$.

\begin{theorem}[\cite{DBLP:conf/innovations/ChenDS14}]
Let $f: \R^\Sigma \to \R$ be a function. Then there is a CRC that $\R_{\geq 0}$-stably computes a dual-rail representation of $f$ if and only if $f$ is continuous and piecewise rational linear.
\end{theorem}

It is natural to wonder about the computational complexity of determining whether or not we have $\vc \reach \vd$ for given continuous states $\vc$ and $\vd$. It is shown in \cite{CaseLS16/ContCRNsCompute} that if $\vc$ and $\vd$ have only rational entries, then this problem can be solved in polynomial time. In contrast, the reachability problem for CRNs using the usual reachability relation for states of Section~\ref{sec:CRNs} is much harder, cf.\ Subsection~\ref{ssec:VASs}.

\section{Implementation: DNA strand displacement}\label{sec:implement_DNAsd}

\begin{figure*}
\begin{center}
\begin{tikzpicture}
[SDhead/.style={thick,-{Straight Barb[left]}},
SDseg/.style={thick},
SDAcomplex/.pic = {
\draw [SDhead, red] (0.93,-0.22)--(0,-0.22)
node [midway, below, text={black}] {$t^*$}
node [very near end, below left, text={black}] {3'};
\node (D1) at (0.48,-0.1) {\texttt{AGAGTT}};
\draw [SDseg, blue] (0.93,-0.22)--(5.3,-0.22)
node [midway, below, text={black}] {$u^*$}
node [pos=0.97, below right, text={black}] {5'};
\node (D3) at (3.1,-0.1) {\texttt{ACGATCGGTACGATTCACAGCATGCTAGG}};
\node (D4) at (3.1, 0.1) {\texttt{TGCTAGCCATGCTAAGTGTCGTACGATCC}};
\draw [SDhead, blue] (0.93,0.22)--(5.3,0.22)
node [midway, above, text={black}] {$u$}
node [pos=0.97, above right, text={black}] {3'}
node [pos=0.05, above left, text={black}] {5'};
},
SDBcomplex/.pic = {
\draw [SDhead, red] (1,-0.025)--(0.5,-0.025)
node [midway, below, text={black}] {$t^*$};
\draw [SDseg, blue] (1,-0.025)--(2,-0.025)
node [midway, below, text={black}] {$u^*$};
\draw [SDhead, blue] (1,0.025)--(2,0.025)
node [midway, above, text={black}] {$u$};
}]
\draw (0,0) pic{SDAcomplex};
\node (p1) at (7,0) {\scalebox{1.7}{$=$}} ;
\draw (8,0) pic{SDBcomplex};
\end{tikzpicture}
\end{center}
\caption{DNA molecule representation.}
\label{fig:DNA_repr}
\end{figure*}
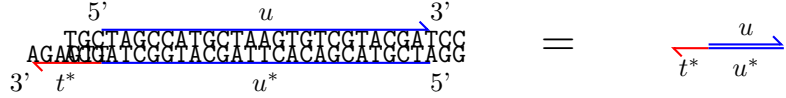

\begin{figure*}
\begin{center}
\begin{tikzpicture}
[SDhead/.style={thick,-{Straight Barb[left]}},
SDseg/.style={thick},
SDAcomplex/.pic = {
\draw [SDseg, red] (0.5,0)--(1,0)
node [midway, above, text={black}] {$t$};
\draw [SDhead, blue] (1,0)--(2,0) 
node [midway, above, text={black}] {$u$};
},
SDBcomplex/.pic = {
\draw [SDhead, red] (1,-0.025)--(0.5,-0.025)
node [midway, below, text={black}] {$t^*$};
\draw [SDseg, blue] (1,-0.025)--(2,-0.025)
node [midway, below, text={black}] {$u^*$};
\draw [SDhead, blue] (1,0.025)--(2,0.025)
node [midway, above, text={black}] {$u$};
},
SDMcomplex/.pic = {
\draw [SDhead, red] (1,-0.025)--(0.5,-0.025)
node [midway, below, text={black}] {$t^*$};
\draw [SDseg, blue] (1,-0.025)--(2,-0.025)
node [midway, below, text={black}] {$u^*$};
\draw [SDhead, blue] (1,0.025)--(2,0.025)
node [midway, above, text={black}] {$u$};
\draw [SDhead, blue] (1,0.025)--(1.7,0.7)
node [near end, left, text={black}] {$u$};
\draw [SDseg, red] (0.5,0.025)--(1,0.025)
node [midway, above, text={black}] {$t$};
},
SDEcomplex/.pic = {
\draw [SDseg, red] (0.5,0.025)--(1,0.025)
node [midway, above, text={black}] {$t$};
\draw [SDhead, red] (1,-0.025)--(0.5,-0.025)
node [midway, below, text={black}] {$t^*$};
\draw [SDseg, blue] (1,-0.025)--(2,-0.025)
node [midway, below, text={black}] {$u^*$};
\draw [SDhead, blue] (1,0.025)--(2,0.025)
node [midway, above, text={black}] {$u$};
},
SDFcomplex/.pic = {
\draw [SDhead, blue] (0,0)--(1,0) 
node [midway, above, text={black}] {$u$};
}
]
\draw (0,1) pic{SDAcomplex};
\draw (0,0) pic{SDBcomplex};
\draw [->, thick] (3,0.3)--(4,0.3);
\draw [<-, thick] (3,0.7)--(4,0.7);
\draw (4.2,0.5) pic{SDMcomplex};
\draw [->, thick] (7,0.5)--(8,0.5);
\draw (9,1) pic{SDFcomplex};
\draw (8.5,0) pic{SDEcomplex};
\end{tikzpicture}
\end{center}
\caption{DNA strand displacement.}
\label{fig:strand_displacement_principle}
\end{figure*}
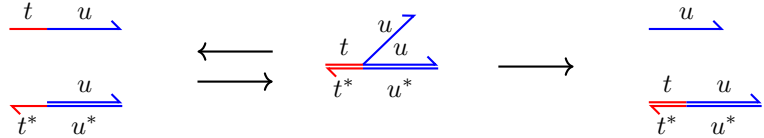

In the previous sections we have seen various ways in which (abstract) CRNs can perform computations. We now discuss from \cite{SimCRN/Soloveichik} a method of implementing an arbitrary (abstract) CRN $\calN$ in the wetlab using DNA as a substrate. 

First, let us use a concise representation of a DNA molecule that abstracts away from the exact identity of the DNA base-pair sequence. The left-hand side of Figure~\ref{fig:DNA_repr} depicts a DNA molecule where one single strand consisting of the segments $u^*$ and $t^*$ is bound to the single strand $u$ complementary to $u^*$ (in general, we denote by $x^*$ the Watson-Crick complement of $x$). As usual, a single strand is denoted by an arrow and its 3'-end is denoted by an arrow head. For visual clarity, we use colors to emphasize the various segments of a single strand/arrow. The concise representation of the left-hand side of Figure~\ref{fig:DNA_repr} is given on the right-hand side of that figure. 

We now discuss the key principle of DNA strand displacement, illustrated in Figure~\ref{fig:strand_displacement_principle}. Since $t$ and $t^*$ are complementary segments appearing on the left-hand side of Figure~\ref{fig:strand_displacement_principle}, these segments can bind, which results in a single DNA molecule given in the middle part of Figure~\ref{fig:strand_displacement_principle}. Segment $t$ is a small segment, called a \emph{toehold}, designed to be small enough for the binding to be reversible. Thus, it may happen that $t$ and $t^*$ unbind and we obtain again the situation on the left-hand side of Figure~\ref{fig:strand_displacement_principle}. Alternatively, the two $u$ segments may compete for binding with $u^*$ in a random walk fashion and it may happen that the segment $u$ that is connected to $t$ completely pushes out the single strand $u$ that was bound to $u^*$ (single strand $u$ is then called \emph{displaced}), see the right-hand side of Figure~\ref{fig:strand_displacement_principle}. Note that this second step of pushing out the single strand $u$ is irreversible.

\newcommand{\SDmol}[2]{
\draw (1.5,0.5) node {\small{Molecule ${#2}$}};
\draw[dotted] (-0.1,-0.4) rectangle (3.1,0.3);
\draw [SDseg, black] (0,-0.2)--(1,-0.2) node [midway, above] {{#1}};
\draw [SDseg, red] (1,-0.2)--(1.5,-0.2) node [midway, above, text={black}] {$i_{#2}$};
\draw [SDseg, blue] (1.5,-0.2)--(2.5,-0.2) node [midway, above, text={black}] {$s_{#2}$};
\draw [SDhead, green] (2.5,-0.2)--(3,-0.2) node [midway, above, text={black}] {$o_{#2}$};
}

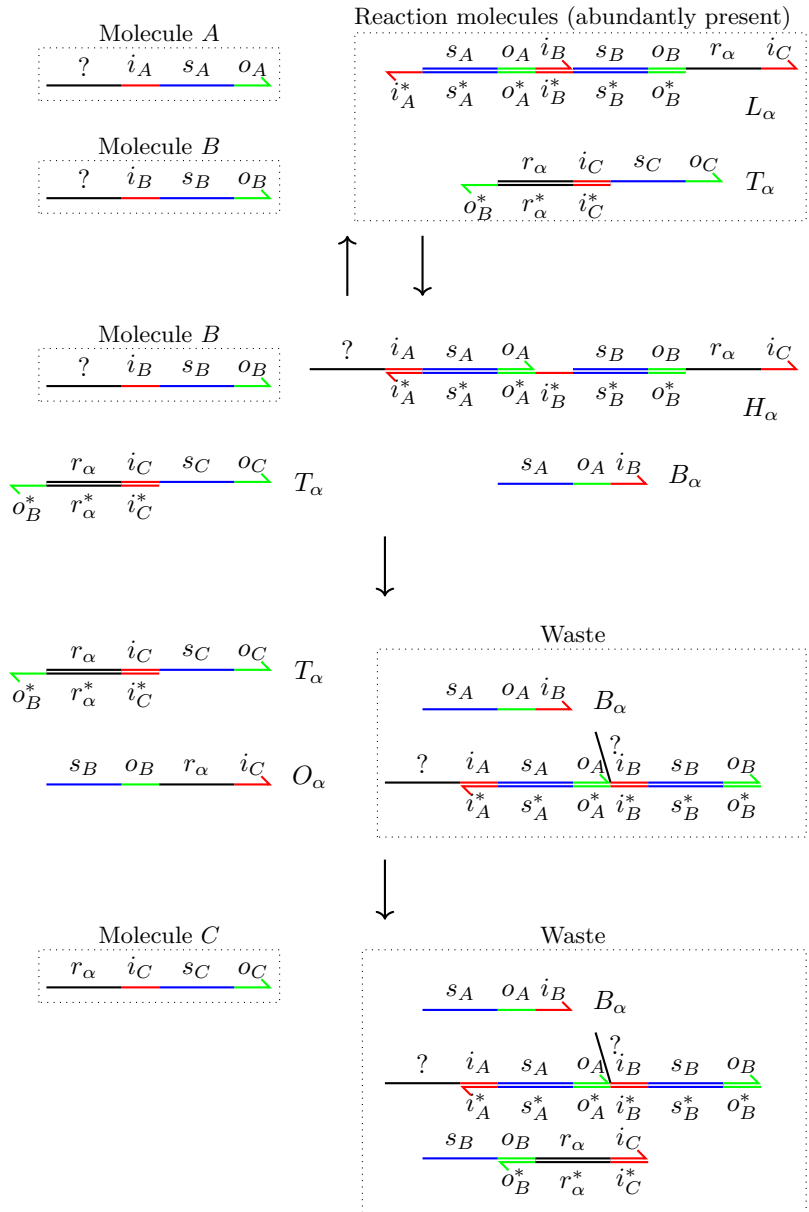
\begin{figure*}
\begin{center}
\begin{tikzpicture}
[SDhead/.style={thick,-{Straight Barb[left]}},
SDseg/.style={thick},
SDRcomplex/.pic = {
\draw [SDhead, red] (1,-0.025)--(0.5,-0.025)
node [midway, below, text={black}] {$i_A^*$};
\draw [SDseg, blue, double] (1,0)--(2,0) 
node [midway, above, text={black}] {$s_A$} 
node [midway, below, text={black}] {$s_A^*$};
\draw [SDseg, green, double] (2,0)--(2.5,0) 
node [midway, above, text={black}] {$o_A$}
node [midway, below, text={black}] {$o_A^*$};
\draw [SDhead, red] (2.5,0.025)--(3,0.025) 
node [midway, above, text={black}] {$i_B$}
node [midway, below, text={black}] {$i_B^*$};
\draw [SDseg, red] (2.5,-0.025)--(3,-0.025);
\draw [SDseg, blue, double] (3,0)--(4,0) 
node [midway, above, text={black}] {$s_B$}
node [midway, below, text={black}] {$s_B^*$};
\draw [SDseg, green, double] (4,0)--(4.5,0) 
node [midway, above, text={black}] {$o_B$}
node [midway, below, text={black}] {$o_B^*$};
\draw [SDseg, black] (4.5,0.025)--(5.5,0.025) 
node [midway, above, text={black}] {$r_\alpha$};
\draw [SDhead, red] (5.5,0.025)--(6,0.025) 
node [midway, above, text={black}] {$i_C$};
\draw (5.5,-0.5) node {$L_\alpha$};
},
SDAcomplex/.pic = {
\SDmol{?}{A}
},
SDBcomplex/.pic = {
\SDmol{?}{B}
},
SDCcomplex/.pic = {
\SDmol{$r_\alpha$}{C}
},
SDRcomplexTwo/.pic = {
\draw [SDhead, green] (0,-0.025)--(-0.5,-0.025)
node [midway, below, text={black}] {$o_B^*$};
\draw [SDseg, black, double] (0,0)--(1,0) 
node [midway, above] {$r_\alpha$}
node [midway, below] {$r_\alpha^*$};
\draw [SDseg, red, double] (1,0)--(1.5,0) 
node [midway, above, text={black}] {$i_C$}
node [midway, below, text={black}] {$i_C^*$};
\draw [SDseg, blue] (1.5,0.025)--(2.5,0.025) 
node [midway, above, text={black}] {$s_C$};
\draw [SDhead, green] (2.5,0.025)--(3,0.025) 
node [midway, above, text={black}] {$o_C$};
\draw (3.5,0) node {$T_\alpha$};
},
SDRcomplexThree/.pic = {
\draw [SDseg, black] (-0.5,0.025)--(0.5,0.025)
node [midway, above, text={black}] {?};
\draw [SDseg, red] (0.5,0.025)--(1,0.025) 
node [midway, above, text={black}] {$i_A$} 
node [midway, below, text={black}] {$i_A^*$};
\draw [SDhead, red] (1,-0.025)--(0.5,-0.025);
\draw [SDseg, blue, double] (1,0)--(2,0) 
node [midway, above, text={black}] {$s_A$} 
node [midway, below, text={black}] {$s_A^*$};
\draw [SDhead, green] (2,0.025)--(2.5,0.025) 
node [midway, above, text={black}] {$o_A$}
node [midway, below, text={black}] {$o_A^*$};
\draw [SDseg, green] (2,-0.025)--(2.5,-0.025);
\draw [SDseg, red] (2.5,-0.025)--(3,-0.025) 
node [midway, below, text={black}] {$i_B^*$};
\draw [SDseg, blue, double] (3,0)--(4,0) 
node [midway, above, text={black}] {$s_B$}
node [midway, below, text={black}] {$s_B^*$};
\draw [SDseg, green, double] (4,0)--(4.5,0) 
node [midway, above, text={black}] {$o_B$}
node [midway, below, text={black}] {$o_B^*$};
\draw [SDseg, black] (4.5,0.025)--(5.5,0.025) 
node [midway, above, text={black}] {$r_\alpha$};
\draw [SDhead, red] (5.5,0.025)--(6,0.025) 
node [midway, above, text={black}] {$i_C$};
\draw (5.5,-0.5) node {$H_\alpha$};
},
SDWcomplex/.pic = {
\draw [SDseg, blue] (0,0)--(1,0) node [midway, above, text={black}] {$s_A$};
\draw [SDseg, green] (1,0)--(1.5,0) node [midway, above, text={black}] {$o_A$};
\draw [SDhead, red] (1.5,0)--(2,0) node [midway, above, text={black}] {$i_B$};
\draw (2.5,0.1) node {$B_\alpha$};
},
SDRcomplexFour/.pic = {
\draw [SDseg, blue] (0,0)--(1,0) 
node [midway, above, text={black}] {$s_B$};
\draw [SDseg, green] (1,0)--(1.5,0) 
node [midway, above, text={black}] {$o_B$};
\draw [SDseg, black] (1.5,0)--(2.5,0) 
node [midway, above, text={black}] {$r_\alpha$};
\draw [SDhead, red] (2.5,0)--(3,0) 
node [midway, above, text={black}] {$i_C$};
\draw (3.5,0.1) node {$O_\alpha$};
},
SDWcomplexTwo/.pic = {
\draw [SDseg, black] (-0.5,0.025)--(0.5,0.025)
node [midway, above, text={black}] {?};
\draw [SDseg, red] (0.5,0.025)--(1,0.025) 
node [midway, above, text={black}] {$i_A$} 
node [midway, below, text={black}] {$i_A^*$};
\draw [SDhead, red] (1,-0.025)--(0.5,-0.025);
\draw [SDseg, blue, double] (1,0)--(2,0) 
node [midway, above, text={black}] {$s_A$} 
node [midway, below, text={black}] {$s_A^*$};
\draw [SDhead, green] (2,0.025)--(2.5,0.025) 
node [midway, above, text={black}] {$o_A$}
node [midway, below, text={black}] {$o_A^*$};
\draw [SDseg, green] (2,-0.025)--(2.5,-0.025);
\draw [SDseg] (2.5,0.025)--(2.3,0.7)
node [near end, right] {?};
\draw [SDseg, red, double] (2.5,0)--(3,0)
node [midway, above, text={black}] {$i_B$}
node [midway, below, text={black}] {$i_B^*$};
\draw [SDseg, blue, double] (3,0)--(4,0) 
node [midway, above, text={black}] {$s_B$}
node [midway, below, text={black}] {$s_B^*$};
\draw [SDhead, green] (4,0.025)--(4.5,0.025) 
node [midway, above, text={black}] {$o_B$}
node [midway, below, text={black}] {$o_B^*$};
\draw [SDseg, green] (4,-0.025)--(4.5,-0.025);
},
SDWcomplexThree/.pic = {
\draw [SDseg, blue] (0,0.025)--(1,0.025) 
node [midway, above, text={black}] {$s_B$};
\draw [SDseg, green] (1,0.025)--(1.5,0.025) 
node [midway, above, text={black}] {$o_B$}
node [midway, below, text={black}] {$o_B^*$};
\draw [SDhead, green] (1.5,-0.025)--(1,-0.025);
\draw [SDseg, black, double] (1.5,0)--(2.5,0) 
node [midway, above, text={black}] {$r_\alpha$}
node [midway, below, text={black}] {$r_\alpha^*$};
\draw [SDhead, red] (2.5,0.025)--(3,0.025) 
node [midway, above, text={black}] {$i_C$}
node [midway, below, text={black}] {$i_C^*$};
\draw [SDseg, red] (2.5,-0.025)--(3,-0.025);
}
]

\draw (0,0) pic{SDAcomplex};

\draw (0,-1.5) pic{SDBcomplex};

\draw (7,0.7) node {\small{Reaction molecules (abundantly present)}};
\draw[dotted] (4.1,-2) rectangle (10.1,0.5);
\draw (4,0) pic{SDRcomplex};

\draw (6,-1.5) pic{SDRcomplexTwo};

\draw [->, thick] (5,-2.2)--(5,-3);
\draw [<-, thick] (4,-2.2)--(4,-3);

\draw (0,-4) pic{SDBcomplex};
\draw (0,-5.5) pic{SDRcomplexTwo};
\draw (4,-4) pic{SDRcomplexThree};
\draw (6,-5.5) pic{SDWcomplex};

\draw [->, thick] (4.5,-6.2)--(4.5,-7);

\draw (0,-8) pic{SDRcomplexTwo};
\draw (0,-9.5) pic{SDRcomplexFour};

\draw (7,-7.5) node {\small{Waste}};
\draw[dotted] (4.4,-10.2) rectangle (10.1,-7.7);
\draw (5,-8.5) pic{SDWcomplex};
\draw (5,-9.5) pic{SDWcomplexTwo};

\draw [->, thick] (4.5,-10.5)--(4.5,-11.3);

\draw (0,-12) pic{SDCcomplex};

\draw (7,-11.5) node {\small{Waste}};
\draw[dotted] (4.2,-15.2) rectangle (10.1,-11.7);
\draw (5,-12.5) pic{SDWcomplex};
\draw (5,-13.5) pic{SDWcomplexTwo};
\draw (5,-14.5) pic{SDWcomplexThree};
\end{tikzpicture}
\end{center}
\caption{Simulating reaction $\alpha = A + B \to C$ through DNA strand displacements.}
\label{fig:strand_displacement}
\end{figure*}

Figure~\ref{fig:strand_displacement} gives now the implementation of an example reaction $\alpha = A + B \to C$ using DNA strand displacement from \cite{SimCRN/Soloveichik}. A molecule of $A$ is represented by a single strand consisting of four segments. The black segment can be arbitrary (although we naturally assume that different segments are always sufficiently different from the (complements of the) other segments of the figure so that they not interfere in unintended ways \cite{Winfree_DNAseq_design,Lorenz_DNAseq_design}), and the segments $i_A$, $s_A$, $o_A$ together form an identifier for species $A$. The segments $i_A$ and $o_A$ are toeholds. Aside from these single strands, there are additional molecules $L_\alpha$ and $T_\alpha$ which are assumed to be abundantly present in the well-mixed solution. 

If an $A$-molecule is present, then the ``incoming toehold'' $i_A$ can bind to its complement $i_A^*$ in molecule $L_\alpha$. Again, because of the small size of $i_A$, the single strand representing $A$ may also unbind at this stage. Alternatively, it may compete with the existing single strand $B_\alpha = s_A o_A i_B$ that is part of $L_\alpha$ and possibly push $B_\alpha$ out obtaining $H_\alpha$. Note that this process is reversible as $B_\alpha$ has $i_B$ as a toehold which can bind to $i_B^*$ in $H_\alpha$ and push out the molecule representing $A$. Alternatively, if a $B$-molecule is present, then it may also bind to $i_B^*$ in $H_\alpha$ and this may result in pushing out single strand $O_\alpha$. The remainder of $H_\alpha$ is waste, and at this stage $B_\alpha$ is waste too. Note that this step is irreversible since $O_\alpha$ cannot bind to the remainder of $H_\alpha$. Finally, $O_\alpha$ can bind to toehold $o_B^*$ of $T_\alpha$ and this may result in pushing out a single strand that represents $C$. Again, this step is irreversible. It is important that the first of the three steps is reversible. Indeed, if no $B$-molecules are present and the first step is irreversible, then $A$-molecules would be incorrectly consumed by the $L_\alpha$-molecules.

Since we consume an $L_\alpha$ and a $T_\alpha$ molecule for every application of the reaction $\alpha$, it is necessary to keep adding these ``fuel'' molecules to the well-mixed solution to ensure reaction $\alpha$ can keep taking place. 

Additional systematic methodologies for compiling a given (abstract) CRN to an implementation are given in \cite{ChenDSPCSS/CRNprog/NatureNano,BadeltSJDTW17/LNCS/Nuskell}. Note that an implementation of an (abstract CRN) is a CRN as well, called an \emph{implementation CRN}. Verification of correctness of an implementation CRN against an abstract CRN has been studied using the notion of pathway decomposition in \cite{ShinTW14/BisimCRNs} and using the notion of bisimulation in \cite{JohnsonDW16/BisimCRNs}. We also remark that the notion of correctness in general depends on the computational model that is assumed. For example, the implementation CRN of Figure~\ref{fig:strand_displacement} would in general not faithfully represent the original abstract CRN if we assume a computational model that highly depends on specific values of the rate constants, like the model of computing with probability density functions in Subsection~\ref{ssec:comp_prob_dist}.

Note that for CRNs $\calN$ having ``non mass-conserving'' reactions like $\vec{0} \to A$ or $A \to \vec{0}$, where $\vec{0}$ is a zero vector, we necessarily need ``fuel'' molecules or ``waste'' molecules, respectively, for any implementation of $\calN$ in nature (notice that the above mentioned implementation of a CRN by DNA strand displacement needs fuel molecules and has waste molecules also for mass-conserving reactions). Mass-conserving CRNs are exactly the CRNs for which it is possible to assign positive integers to the species, a weight vector $\vec{v}$, such that the weighted sum of the molecules in a state is invariant under the application of any reaction (i.e., $\vec{v}^T M = \vec{0}^T$, where $M$ is the stoichiometry matrix of the CRN). Such a weight vector is called a \emph{conservation vector} \cite{Horn1972/CRNTheory/ConservVector}.

\section{Related research fields}\label{sec:related_fields}

Since CRNs form a mathematically natural model, it is not surprising that this notion (or notions very similar to it) has also appeared in other contexts. Indeed, CRNs are very closely related to the notions of Petri nets \cite{PetriNet/review/Pet1977,DBLP:conf/ac/1996petri1} and vector addition systems \cite{VASsKarpMiller} from the theory of concurrency and population protocols \cite{DBLP:journals/eatcs/AspnesR07} from the theory of distributed computing.

\subsection{Petri nets}

We first turn to Petri nets \cite{PetriNet/review/Pet1977,DBLP:conf/ac/1996petri1}, which are nearly identical to CRNs. In a Petri net, molecules are called \emph{tokens}, species are called \emph{places}, reactions are called \emph{transitions}, and states are called \emph{markings}. The \emph{firing} of a transition in a Petri net corresponds to a reaction that takes place in a CRN\@. More advanced notions often also have their counterpart in Petri net theory (and vice versa), e.g., the notion of a conservation vector (mentioned in Section~\ref{sec:implement_DNAsd}) is called a \emph{P-invariant} in Petri net theory.

A Petri net is defined as a directed bipartite multigraph where the vertices of one colour class $P$ are the places (and are depicted as round vertices) and the vertices of the other colour class $T$ are the transitions (and are depicted as square vertices).

\begin{figure}
\begin{center}
\tikzstyle{place}=[circle,draw=blue!50,fill=blue!20,thick,
inner sep=0pt,minimum size=6mm]
\tikzstyle{transition}=[rectangle,draw=black!50,fill=black!20,thick,
inner sep=0pt,minimum size=4mm]
\begin{tikzpicture}
\node[place] (x1) {$X_1$};
\node[place] (x2) [below of=x1] {$X_2$};
\node[place] (x3) [below of=x2] {$X_3$};
\node[transition] (a) [left of=x2,xshift=-5mm] {$\alpha$};
\node[transition] (b) [right of=x2,xshift=5mm] {$\beta$};
\draw [-Latex] (x2) to [bend right=10] (a);
\draw [-Latex] (x2) to [bend left=10] (a);
\draw [-Latex] (x1) to [bend right=40] (a);
\draw [-Latex] (a) to [bend right=10] (x1);
\draw [-Latex] (a) to [bend left=10] (x1);
\draw [-Latex] (a) to [bend right=40] (x3);
\draw [-Latex] (x3) to [bend right=40] (b);
\draw [-Latex] (b) to [bend right=40] (x1);
\end{tikzpicture}
\end{center}
\caption{The Petri net corresponding to the CRN $\calN$ of Example~\ref{ex:CRN_Petri}.}
\label{fig:ex_Petrinet}
\end{figure}
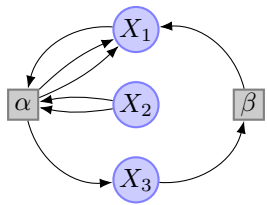

\begin{example}\label{ex:CRN_Petri}
Consider the CRN $\calN = (\{X_1,X_2,X_3\},\allowbreak\{\alpha,\beta\})$ with
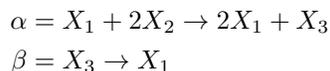
\begin{align*}
\alpha &= X_1 + 2X_2 \to 2X_1 + X_3\\
\beta &= X_3 \to X_1
\end{align*}
The Petri net corresponding to $\calN$ is depicted in Figure~\ref{fig:ex_Petrinet}. Note that the reactions correspond to square vertices (i.e., transitions) and that the species correspond to round vertices (i.e., places) in Figure~\ref{fig:ex_Petrinet}. The reactants (products, resp.) of each reaction correspond to the incoming (outgoing, resp.) arrows, with multiplicity, of the corresponding transition. For example since, $X_1+2X_2$ are the reactants of $\alpha$, there is one arrow from $X_1$ to $\alpha$ and two arrows from $X_2$ to $\alpha$.
\end{example}

There are some small differences between the (usual) definitions of a CRN and a Petri net, which for most problems are irrelevant. One difference is that a Petri net has an initial marking (i.e., an initial state), while this is not the case in the (usual) definition of a CRN\@. Of course, such a fixed initial state can be useful in the context of CRNs as well (see, e.g., Subsection~\ref{ssec:comp_prob_dist}). A more subtle difference is that a Petri net may have two (or more) transitions with the same multisets of incoming and outgoing arrows, which would corresponds to two distinct reactions of the form $(\vr,\vp)$.

Many results on Petri nets deal with behavioral properties starting from the initial marking/state. Roughly speaking, a Petri net is most often considered as a \emph{generator} of states. In contrast, the computational CRN models of Section~\ref{sec:comp_CRNs} deal with accepting or rejecting an unknown input state. The models from Section~\ref{sec:comp_CRNs} have not been considered in the context of Petri net theory but have been taken from the theory of population protocols, cf.\ Subsection~\ref{ssec:PopProt} below. 

The notion of a \emph{stochastic Petri net} \cite{DBLP:conf/eef/Balbo00,StochPetriNetsBook,DBLP:conf/apn/Marsan88} studied in the literature is similar to the notion of a stochastic CRN, however the notion of propensity (defined in Subsection~\ref{ssec:stoch_CRNs}) that is used is different. More specifically, in stochastic Petri nets the propensity is \emph{equal} to the rate constant (called \emph{firing rate} in the context of stochastic Petri nets) and so, e.g., transition $t$ can fire for markings $\vc$ and $\vd$, then the expected time for $t$ to fire for $\vc$ is equal to the expected time for $t$ to fire for $\vd$ --- this behavior is, of course, very different from stochastic CRNs (indeed, for stochastic CRNs the expected time for a reaction to take place decreases when increasing the molecule counts of species that appear as reactants of that reaction).

The notion of a \emph{continuous Petri net} \cite{DBLP:conf/apn/RecaldeTS99,HybridContinuousPetriBook} studied in the literature is quite different from the notion of a continuous CRN from Subsection~\ref{ssec:ContinuousCRNs}. Indeed, the former is not based on differential equations, but instead allows transitions/reactions to be applied ``$x \in \R$ times''. In this way, continuous Petri nets are more related to the rate-independent mode of operation discussed in Subsection~\ref{ssec:rateIndepCRNcomp}.

We mention that there are other classes of Petri nets, like hybrid Petri nets \cite{HybridContinuousPetriBook} and coloured Petri nets \cite{ColouredPetriNetsBook}, which currently have not yet been considered in the context of CRNs.

\subsection{Vector addition systems}\label{ssec:VASs}
A \emph{vector addition system} (\emph{VAS} for short) \cite{VASsKarpMiller} is a finite subset $A$ of $\Z^\Lambda$, where $\Z$ is the set of integers and $\Lambda$ is finite. The elements of $A$ are called \emph{actions}. Similar as for CRNs and Petri nets, a \emph{state} is an element of $\N^\Lambda$. An action $\vec{a} \in A$ can \emph{fire} for state $\vc$ if $\vc+\vec{a}$ is a state. 

For an action $\vec{a}$, let $\vp, \vr \in \N^\Lambda$ be such that for all $X \in \Lambda$, we have (1) $\vp(X) = \vec{a}(X)$ and $\vr(X) = 0$ if $\vec{a}(X) \geq 0$ and (2) $\vr(X) = -\vec{a}(X)$ and $\vp(X) = 0$ otherwise. Then $\vec{a} = \vp - \vr$, and we can easily see that the reaction $(\vr,\vp)$ simulates the action $\vec{a}$. In this way, for each VAS $A$ there is a CRN that simulates $A$.

There is, however, an issue in simulating a CRN by a VAS. Consider the VAS $A$ obtained from a CRN $\calN$ by replacing each reaction $(\vr,\vp)$ by the action $\vp - \vr$. Then $A$ may behave \emph{differently} than $\calN$. Indeed, for example, the reaction $A+3B \to C+3B$ cannot be applied to a state $\vd$ with only the single molecule $A$, but the action $\vp - \vr = C-A$ \emph{can} be applied to $\vd$. More generally, the construction does not work for ``catalyst-like'' reactions $(\vr,\vp)$, where $\vr(X)$ and $\vp(X)$ are both nonzero for some species $X$. However, for each CRN $\calN$ there is a CRN $\calN'$ without catalyst-like reactions that behaves very similar to $\calN$. The CRN $\calN'$ is obtained from $\calN$ by introducing a new species $Q_\alpha$ for each catalyst-like reaction $\alpha = (\vr,\vp)$ and replacing $\alpha$ by the reactions $\alpha_1 = (\vr,\vec{q}_\alpha)$ and $\alpha_2 = (\vec{q}_\alpha,\vp)$, where $\vec{q}_\alpha$ contains one copy of $Q_\alpha$ and nothing else \cite{DBLP:journals/nc/SoloveichikCWB08,CookSWB2009/CRNSurvey}. In this way, e.g., the reaction $\alpha = A+3B \to C+3B$ is simulated by the reactions $A+3B \to Q_\alpha$ and $Q_\alpha \to C+3B$. For many problems the difference between $\calN$ and $\calN'$ is irrelevant and for these problems we can equivalently consider the VAS corresponding to $\calN'$.

The reachability problem for VASs, i.e., to determine for given states $\vc$ and $\vd$ whether or not $\vd$ can be reached from $\vc$, has been intensively investigated and is well known to be EXPSPACE-hard \cite{DBLP:conf/stoc/CardozaLM76} (lower bound) and decidable \cite{VASsMayr/siamcomp,Leroux/VASs/simpleproof} (upper bound, see the introduction of \cite{Leroux/VASs/simpleproof} for a more detailed historical account of the decidability proofs). By the above, these results directly carry over to the domains of Petri nets and CRNs.

\subsection{Population protocols}\label{ssec:PopProt}

The notion of a \emph{population protocol} was introduced in \cite{DBLP:journals/dc/AngluinADFP06} as a model for distributed computing. A population protocol models a finite set of \emph{agents} that each hold a \emph{state} from a fixed finite set $Q$ of states. When two agents bump into each other, the agents change their state according to a transition function $\delta: Q^2 \to Q^2$. Agents with a common state are indistinguishable, so a particular global state of a set of agents can be described as a multiset of the states of the agents. We can now easily see that population protocols correspond to CRNs where each reaction $(\vr,\vp)$ is such that $\|\vr\| = \|\vp\| = 2$. Indeed, agents correspond to molecules, states correspond to species, and if $\delta(q_1,q_2) = (q_3,q_4)$, then this corresponds to reaction $q_1 + q_2 \to q_3 + q_4$. More precisely, the class of population protocols therefore actually corresponds to the class of CRNs where each reaction $(\vr,\vp)$ is such that $\|\vr\| = \|\vp\| = 2$ \emph{and}, additionally, there is a reaction for each pair of species. However, we may have $\delta(q_1,q_2) = (q_1,q_2)$ and so the corresponding reaction $(\vr,\vp)$ is mute. For most problems the existence or absence of mute reactions is irrelevant. The computational CRN model of Subsection~\ref{ssec:stably_deciding} is the natural generalization to CRNs of the original computational model for population protocols from \cite{DBLP:journals/dc/AngluinADFP06}. The interpretation of the computational model of Subsection~\ref{ssec:stably_deciding} in terms of population protocols is as follows: each agent starts with an input state and ``eventually'' there is agreement among the agents of accepting the input or not (we use eventually in the sloppy way here, cf.\ Remark~\ref{rem:eventually}: we actually mean ``during the computation it is always possible to reach a state where''). Some states are designated as ``yes'' states and others as ``no'' states --- in this way, agents communicate their opinion. Notice that the notion of stably deciding is natural within the context of population protocols since agents will keep bumping into each other, triggering the application of the transition function $\delta$. Indeed, stably deciding (not haltingly deciding) is the most studied mode of operation for population protocols.

Because the class of population protocols corresponds to a proper subclass of all CRNs, results concerning population protocols do not necessarily hold for the whole class of CRNs. One particularly important aspect of population protocols is that the number of agents stay fixed during a computation. In other words, in the corresponding CRN $\calN$, we have that for all states $\vc$ and $\vd \in \post(\vc)$, $\|\vd\| = \|\vc\|$. Since there are only a finite number of states of a given size, we have that the set $\post(\vc)$ is finite. It is easy to define CRNs that violate this property: take, e.g., a CRN having the reaction $\vec{0} \to A$, where $\vec{0}$ is a zero vector.

The efficiency of population protocol algorithms is expressed in terms of the expected number of interactions between agents, where the two agents for each interaction are chosen at random. While the model is therefore similar to that of stochastic CRNs where the rate constants are all $1$ and the volume is equal to the number of agents, there is a difference in that population protocols use discrete time and stochastic CRNs use continuous time.

\section{Discussion}
The goal of this paper is to introduce in a tutorial fashion the basic concepts and results concerning computational CRNs as well as to review some of the main strands of research in this area. 
By now the literature of this research field is really vast, so it is not possible to cover (in a space-limited tutorial) all interesting research directions. We complete this tutorial by mentioning a few research directions that we did not cover.

Most prominently, we have not discussed the important topic of model checking, i.e., verifying behavioral properties of CRNs. CRN theory, see, e.g., \cite{Feinberg/CRNLectures,FeinbergHorn/deficiency,Horn/deficiency,Gunawardena/CRNSurvey}, is a well-established research field that is traditionally used to study CRNs occurring in nature, but it can equally well be used to model check human-designed computational CRNs. Model checking techniques can be drawn from various contexts. Indeed, for example, various notions introduced originally in the context of continuous CRNs, such as the important notion of deficiency, have found their use also for discrete CRNs \cite{Anderson/CRN/Domination}. As another example, notions introduced originally in the context of Petri nets, such as the notion of a T-invariant, have found their use for discrete CRNs, see, e.g., \cite{Brijder17/NaCo/Tinvariant}.

Also, since CRNs behave inherently asynchronously, it is natural to link CRNs to asynchronous logic circuits. 
This research direction is pursued in \cite{CardelliKW16/CRN/AsyncLogicCircuits} where (among other results) it is shown that the Muller C-element (a fundamental asynchronous component) \cite{Book/AsynchronousCircuits} can be simulated by a CRN\@. Other work on asynchronous logic circuits implemented by CRNs includes \cite{CRN_asynchronous}.

CRNs have a finite number of reactions and a finite number of species. It would be interesting to see what results concerning CRNs hold in the more general setting where we drop one or both of these assumptions (of course, allowing an infinite number of species without allowing an infinite number of reactions makes little sense). Motivated by polymers, which can be of arbitrary length, a special class of CRNs with an infinite number of species has been considered in \cite{JohnsonWinfree/CRNPolymer}.

In this paper we have also assumed that CRNs reside in well-mixed solutions. However, one can also consider non-homogeneous environments. For example, in \cite{QianW14/DNA20/SurfaceCRN} it is shown possible to implement CRNs tied to surfaces, which are called \emph{surface CRNs}. The ``spatial awareness'' of surface CRNs results in a higher computational expressivity compared to (the usual) CRNs that reside in well-mixed solutions. Indeed, surface CRNs can simulate arbitrary Turing machines (without any theoretical probability of error) \cite{QianW14/DNA20/SurfaceCRN}.

\ifjournal
\begin{acknowledgements}
\else
\subsection*{Acknowledgments}
\fi
We thank Dave Doty, Grzegorz Rozenberg, David Soloveichik, and three anonymous referees for many useful comments on earlier versions of this paper. R.B.\ is a postdoctoral fellow of the Research Foundation -- Flanders (FWO).
\ifjournal
\end{acknowledgements}
\fi

\ifbiblatex
\begingroup
\setlength{\emergencystretch}{8em}
\printbibliography
\endgroup
\else
\bibliography{../crns}
\fi

\end{document}